\documentclass[12pt,a4paper]{article}
\makeatletter
\def\eqnarray{%
\stepcounter{equation}%
\let\@currentlabel=\theequation
\global\@eqnswtrue
\global\@eqcnt\z@
\tabskip\@centering
\let\\=\@eqncr
$$\halign to \displaywidth\bgroup\@eqnsel\hskip\@centering
$\displaystyle\tabskip\z@{##}$&\global\@eqcnt\@ne
\hfil$\displaystyle{{}##{}}$\hfil
&\global\@eqcnt\tw@$\displaystyle\tabskip\z@{##}$\hfil
\tabskip\@centering&\llap{##}\tabskip\z@\cr}
\makeatother
\newcommand{\ket}[1]{{\vert{#1}\rangle}}

\newcommand{\fukuso}{{\mathbf C}}

\begin{document}

\title{\sl Cavity QED and Quantum Computation in the Weak Coupling Regime II 
: Complete Construction of the Controlled--Controlled NOT Gate}
\author{
  Kazuyuki FUJII 
  \thanks{E-mail address : fujii@yokohama-cu.ac.jp },
  Kyoko HIGASHIDA 
  \thanks{E-mail address : s035577d@yokohama-cu.ac.jp },
  Ryosuke KATO 
  \thanks{E-mail address : s035559g@yokohama-cu.ac.jp }, 
  Yukako WADA 
  \thanks{E-mail address : s035588a@yokohama-cu.ac.jp }\\
  Department of Mathematical Sciences\\
  Yokohama City University\\
  Yokohama, 236--0027\\
  Japan
  }
\date{}
\maketitle
%
%
%
%
\begin{abstract}
  In this paper we treat a cavity QED quantum computation. Namely, 
  we consider a model of quantum computation based on n atoms of 
  laser--cooled and trapped linearly in a cavity and realize it as the n atoms 
  Tavis--Cummings Hamiltonian interacting with n external (laser) fields. 
  
  We solve the Schr{\" o}dinger equation of the model in the weak coupling 
  regime to construct the controlled NOT gate in the case of n=2, and to 
  construct the controlled--controlled NOT gate in the case of n=3 by making 
  use of several resonance conditions and rotating wave approximation 
  associated to them. We also present an idea to construct general quantum 
  circuits.
  
  The approach is more sophisticated than that of the paper [K. Fujii, 
  Higashida, Kato and Wada, Cavity QED and Quantum Computation in the Weak 
  Coupling Regime, J. Opt. B : Quantum Semiclass. Opt. {\bf 6} (2004), 502].
  
  Our method is not heuristic but completely mathematical, and the significant 
  feature is based on a consistent use of Rabi oscillations. 
\end{abstract}
%

%
%
%
%

\section{Introduction}

Quantum Computation (or Computer) is a challenging task in this century 
for not only physicists but also mathematicians. 
Quantum Computation is in a usual understanding based on qubits which are 
based on two level systems (two energy levels or fundamental spins) of atoms, 
See \cite{Books} as for general theory of two level systems. 
The essence of Quantum Computation is to construct an element of huge unitary 
group $U(2^{n})$ by manipulating $n$ atoms with photons, laser fields, etc.

In a realistic image of Quantum Computer we need at least one hundred atoms. 
However, we may meet a very severe problem called Decoherence which 
destroys a superposition of quantum states in the process of unitary evolution 
of our system through some influence arising from natural environment. 
At the moment it is not easy to control the decoherence. See for example 
the papers in \cite{decoherence} as an introduction to the problem.  

An optical system like Cavity QED may have some advantage on this problem, 
so we want to consider a model of quantum computation based on $n$ atoms 
of laser--cooled and trapped linearly in a cavity. 
As an approximate model we realize it as the n atoms Tavis--Cummings 
Hamiltonian interacting with n external (laser) fields. As to the 
Tavis--Cummings model see \cite{Papers-1}. 
To perform the quantum computation we must first of all show that our system 
is universal \cite{nine}. 
To show it we must construct the controlled NOT operator (gate) explicitly in 
the case of $n=2$, \cite{nine}, \cite{KF1}.

For that we must embed a system of two--qubits into a space of wave functions 
of the model and solve the Schr{\" o}dinger equation. In a reduced system 
we can construct the controlled NOT by use of some resonance condition and 
the rotating wave approximation associated to it. 
Then we need to assume that the coupling constants are small enough (the weak 
coupling regime in the title). 

Next we want to construct the controlled--controlled NOT operator in the case 
of $n=3$. 
For that purpose the construction of controlled NOT gates of three types is 
required \footnote{In the study of Cavity QED Quantum Computation this 
important point is missed} 
because three atoms are trapped {\bf linearly} in the cavity. 
We have given an idea to construct them in \cite{FHKW}, which is however 
not complete. In this paper we give a complete construction to 
the controlled--controlled NOT gate. 

However, to push on with our method (namely, for the case of $n$ atoms 
($n \geq 5$)) is not easy by some severe technical reason. Therefore 
we present the idea in \cite{FHKW} once more to construct general quantum 
circuits, which will give a general quantum computation.

The contents of this paper are as follows :
\begin{flushleft}
{\bf Section 1}\ \ Introduction \\
{\bf Section 2}\ \ A Model Based on Cavity QED \\
{\bf Section 3}\ \ Quantum Computation \\
\quad \quad \quad {\bf 3.1}\ \ Controlled NOT Gate \\
\quad \quad \quad {\bf 3.2}\ \ Controlled--Controlled NOT Gate \\
{\bf Section 4}\ \ Further Problem \\
{\bf Section 5}\ \ Discussion 
\end{flushleft}

\section{A Model Based on Cavity QED}

We consider a quantum computation model based on n atoms of laser--cooled 
and trapped linearly in a cavity and realize it as the n atoms 
Tavis--Cummings Hamiltonian interacting with n external (laser) fields. 
This is of course an approximate theory. In a more realistic model we must 
add other dynamical variables such as positions of atoms and their momenta 
etc. However, since such a model is almost impossible to solve we consider 
a simple one.

Then the Hamiltonian is given by
\begin{eqnarray}
\label{eq:hamiltonian-1}
H
&=&\omega {1}_{L}\otimes a^{\dagger}a + 
\frac{\Delta}{2} \sum_{j=1}^{n}\sigma^{(3)}_{j}\otimes {\bf 1} +
g\sum_{j=1}^{n}
\left(\sigma^{(+)}_{j}\otimes a+\sigma^{(-)}_{j}\otimes a^{\dagger} \right)+
\nonumber \\
&&\sum_{j=1}^{n}h_{j}
\left(\sigma^{(+)}_{j}\mbox{e}^{i(\Omega_{j}t+\phi_{j})}+
\sigma^{(-)}_{j}\mbox{e}^{-i(\Omega_{j}t+\phi_{j})} \right)\otimes {\bf 1}
\end{eqnarray}
where $\omega$ is the frequency of radiation field, $\Delta$ the energy 
difference of two level atoms, $a$ and $a^{\dagger}$ are 
annihilation and creation operators of the field, and $g$ a coupling constant, 
$\Omega_{j}$ the frequencies of external fields which are treated as classical 
fields, $h_{j}$ coupling constants, and $L=2^{n}$. 
Here $\sigma^{(+)}_{j}$, $\sigma^{(-)}_{j}$ and $\sigma^{(3)}_{j}$ are given 
as 
\begin{equation}
\sigma^{(s)}_{j}=
1_{2}\otimes \cdots \otimes 1_{2}\otimes \sigma_{s}\otimes 1_{2}\otimes \cdots 
\otimes 1_{2}\ (j-\mbox{position})\ \in \ M(L,\fukuso)
\end{equation}
where $s$ is $+$, $-$ and $3$ respectively and 
\begin{equation}
\label{eq:sigmas}
\sigma_{+}=
\left(
  \begin{array}{cc}
    0& 1 \\
    0& 0
  \end{array}
\right), \quad 
\sigma_{-}=
\left(
  \begin{array}{cc}
    0& 0 \\
    1& 0
  \end{array}
\right), \quad 
\sigma_{3}=
\left(
  \begin{array}{cc}
    1& 0  \\
    0& -1
  \end{array}
\right), \quad 
1_{2}=
\left(
  \begin{array}{cc}
    1& 0  \\
    0& 1
  \end{array}
\right).
\end{equation}

\par \noindent
See the figure 1 as an image of the model.
Note that the cases of $n=2$ and $3$ are our target through this paper. 
Here we state our scenario of quantum computation. 
Each independent external field generates a unitary element of the 
corresponding qubit (atom) like 
$a_{1}\otimes a_{2}\otimes \cdots \otimes a_{n}$ where $a_{j}\in U(2)$, 
while a photon inserted generates an entanglement among such elements like 
$\sum a_{1}\otimes a_{2}\otimes \cdots \otimes a_{n}$. 
As a whole we obtain any element in $U(2^{n})$ (a universality).

\vspace{10mm}
\begin{figure}
\begin{center}
\setlength{\unitlength}{1mm} 
\begin{picture}(110,40)(0,-20)
\bezier{200}(20,0)(10,10)(20,20)
\put(20,0){\line(0,1){20}}
\put(30,10){\circle*{3}}
\bezier{200}(30,-4)(32,-2)(30,0)
\bezier{200}(30,0)(28,2)(30,4)
\put(30,4){\line(0,1){2}}
\put(28.6,4){$\wedge$}
\put(40,10){\circle*{3}}
\bezier{200}(40,-4)(42,-2)(40,0)
\bezier{200}(40,0)(38,2)(40,4)
\put(40,4){\line(0,1){2}}
\put(38.6,4){$\wedge$}
\put(50,10){\circle*{1}}
\put(60,10){\circle*{1}}
\put(70,10){\circle*{1}}
\put(50,1){\circle*{1}}
\put(60,1){\circle*{1}}
\put(70,1){\circle*{1}}
\put(80,10){\circle*{3}}
\bezier{200}(80,-4)(82,-2)(80,0)
\bezier{200}(80,0)(78,2)(80,4)
\put(80,4){\line(0,1){2}}
\put(78.6,4){$\wedge$}
\bezier{200}(90,0)(100,10)(90,20)
\put(90,0){\line(0,1){20}}
\put(10,10){\dashbox(90,0)}
\put(99,9){$>$}
\end{picture}
\vspace{-15mm}
\caption{The general setting for a quantum computation based on Cavity QED. 
The dotted line means a single photon inserted in the cavity and all curves 
mean external (laser) fields (which are treated as classical ones) 
subjected to atoms}
\end{center}
\end{figure}
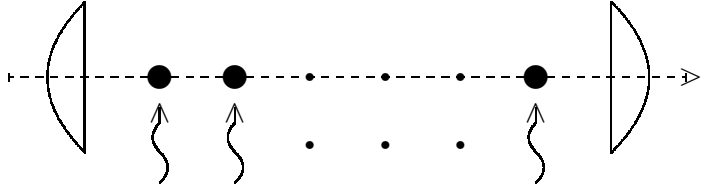
\vspace{-10mm}

Here let us rewrite the Hamiltonian (\ref{eq:hamiltonian-1}). If we set 
\begin{equation}
\label{eq:large-s}
S_{+}=\sum_{j=1}^{n}\sigma^{(+)}_{j},\quad 
S_{-}=\sum_{j=1}^{n}\sigma^{(-)}_{j},\quad 
S_{3}=\frac{1}{2}\sum_{j=1}^{n}\sigma^{(3)}_{j},
\end{equation}
then (\ref{eq:hamiltonian-1}) can be written as 
\begin{eqnarray}
\label{eq:hamiltonian-2}
H
&=&
\omega {1}_{L}\otimes a^{\dagger}a + \Delta S_{3}\otimes {\bf 1}+ 
g\left(S_{+}\otimes a + S_{-}\otimes a^{\dagger} \right)+
\nonumber \\
&&\sum_{j=1}^{n}h_{j}
\left(\sigma^{(+)}_{j}\mbox{e}^{i(\Omega_{j}t+\phi_{j})}+
\sigma^{(-)}_{j}\mbox{e}^{-i(\Omega_{j}t+\phi_{j})} \right)\otimes {\bf 1}
\equiv H_{0}+V(t),
\end{eqnarray}
which is relatively clear. $H_{0}$ is the Tavis--Cummings Hamiltonian and we 
treat it as an unperturved one. 
We note that $\{S_{+},S_{-},S_{3}\}$ satisfy the $su(2)$--relation 
\begin{equation}
[S_{3},S_{+}]=S_{+},\quad [S_{3},S_{-}]=-S_{-},\quad [S_{+},S_{-}]=2S_{3}.
\end{equation}
However, the representation $\rho$ defined by 
\[
\rho(\sigma_{+})=S_{+},\quad \rho(\sigma_{-})=S_{-},\quad 
\rho(\sigma_{3}/2)=S_{3}
\]
is a full representation of $su(2)$, which is of course not irreducible. 

We would like to solve the Schr{\" o}dinger equation 
\begin{equation}
\label{eq:schrodinger}
i\frac{d}{dt}U=HU=\left(H_{0}+V\right)U, 
\end{equation}
where $U$ is a unitary operator. As an equivalent form let us change to the 
interaction picture, which is performed by the 
{\bf method of constant variation}. 
The equation $i\frac{d}{dt}U=H_{0}U$ is solved to be 
\[
U(t)=\left(\mbox{e}^{-it\omega S_{3}}\otimes \mbox{e}^{-it\omega N}\right)
\mbox{e}^{-itg\left(S_{+}\otimes a + S_{-}\otimes a^{\dagger}\right)}
U_{0}
\]
where $N=a^{\dagger}a$ is the number operator and $U_{0}$ a constant unitary. 
Here we have used the resonance condition 
\begin{equation}
\label{eq:old-resonance}
\omega=\Delta
\end{equation}
, see for example \cite{Papers-2}. 
By changing $U_{0}$ $\longmapsto$ $U_{0}(t)$ and substituting into 
(\ref{eq:schrodinger}) we obtain the equation 
\begin{equation}
\label{eq:reduction-equation}
i\frac{d}{dt}U_{0}=
\mbox{e}^{itg\left(S_{+}\otimes a + S_{-}\otimes a^{\dagger}\right)}
\left(\mbox{e}^{it\omega S_{3}}\otimes \mbox{e}^{it\omega N}\right)
V(t)
\left(\mbox{e}^{-it\omega S_{3}}\otimes \mbox{e}^{-it\omega N}\right)
\mbox{e}^{-itg\left(S_{+}\otimes a + S_{-}\otimes a^{\dagger}\right)}
U_{0}
\end{equation}
after some algebras. This is the interaction picture of (\ref{eq:schrodinger}) 
and we use this for our quantum computation.
Therefore we must calculate the right hand side of 
(\ref{eq:reduction-equation}) explicitly, which is however a very hard task 
due to the (complicated) term 
$\mbox{e}^{-itg\left(S_{+}\otimes a + S_{-}\otimes a^{\dagger}\right)}$. 
For convenience in the following we set 
\begin{equation}
\label{eq:A}
A\equiv A_{n}=S_{+}\otimes a + S_{-}\otimes a^{\dagger}.
\end{equation}
It has been done only for $n=1$, $2$, $3$ and $4$ as far as we know, see 
\cite{Papers-2}, \cite{Papers-3}. 
We list the calculations for $n=1$, $2$, $3$ in the following, 
\cite{Papers-2}.

\vspace{3mm}
\par \noindent 
{\bf One Atom Case}\quad In this case $A$ in (\ref{eq:A}) is written as 
\begin{equation}
\label{eq:A-one}
A_{1}=
\left(
  \begin{array}{cc}
    0&           a \\
    a^{\dagger}& 0
  \end{array}
\right)
\equiv B_{1/2}.
\end{equation}
By making use of the simple relation 
\begin{equation}
\label{eq:relation-one}
A_{1}^{2}=
\left(
  \begin{array}{cc}
    aa^{\dagger}&   0          \\
    0           & a^{\dagger}a 
  \end{array}
\right)=
\left(
  \begin{array}{cc}
    N+1& 0  \\
    0  & N
  \end{array}
\right)
\end{equation}
we have 
\begin{eqnarray}
\label{eq:exponential-1/2}
\mbox{e}^{-itgB_{1/2}}=\mbox{e}^{-itgA_{1}}
&=&
\sum_{n=0}^{\infty}\frac{(-1)^n}{(2n)!}
 \left(tg\right)^{2n}A_{1}^{2n}
-i\sum_{n=0}^{\infty}\frac{(-1)^n}{(2n+1)!}
 \left(tg\right)^{2n+1}A_{1}^{2n+1}             \nonumber \\
&=&
\left(
  \begin{array}{cc}
  \mbox{cos}\left(tg\sqrt{N+1}\right)& 
  -i\frac{\mbox{sin}\left(tg\sqrt{N+1}\right)}{\sqrt{N+1}}a  \\
  -i\frac{\mbox{sin}\left(tg\sqrt{N}\right)}{\sqrt{N}}a^{\dagger}& 
  \mbox{cos}\left(tg\sqrt{N}\right)
  \end{array}
\right)            \nonumber \\
&\equiv&
\left(
  \begin{array}{cc}
   C(N+1)            & -iS(N+1)a \\
   -iS(N)a^{\dagger} & C(N)
  \end{array}
\right).
\end{eqnarray}
We obtained the explicit form of solution. However, this form is more or less 
well--known, see for example the second book in \cite{Books}.

\vspace{3mm}
\par \noindent 
{\bf Two Atoms Case}\quad In this case $A$ in (\ref{eq:A}) is written as 
\begin{equation}
\label{eq:A-two}
A_{2}=
\left(
  \begin{array}{cccc}
    0 &          a & a &           0  \\
    a^{\dagger}& 0 & 0 &           a  \\
    a^{\dagger}& 0 & 0 &           a  \\
    0 & a^{\dagger}& a^{\dagger} & 0
  \end{array}
\right).
\end{equation}

Our method is to reduce the $4\times 4$--matrix $A_{2}$ in (\ref{eq:A-two}) to 
a $3\times 3$--matrix $B_{1}$ in the following to make our calculation 
easier. 
For that aim we prepare the following matrix
\begin{equation}
\label{eq:T-two}
T=
\left(
  \begin{array}{cccc}
    0 &   1                & 0                  & 0   \\
    \frac{1}{\sqrt{2}} & 0 & \frac{1}{\sqrt{2}} & 0   \\
   -\frac{1}{\sqrt{2}} & 0 & \frac{1}{\sqrt{2}} & 0   \\
    0 &   0                &   0                  & 1
  \end{array}
\right),
\end{equation}
then it is easy to see 
\[
T^{\dagger}A_{2}T=
\left(
  \begin{array}{cccc}
    0  &                     &                     &            \\
       & 0                   & \sqrt{2}a           & 0          \\
       & \sqrt{2}a^{\dagger} & 0                   & \sqrt{2}a  \\
       & 0                   & \sqrt{2}a^{\dagger} & 0
  \end{array}
\right)\equiv 
\left(
  \begin{array}{cc}
     0 &       \\
       & B_{1} 
  \end{array}
\right)
\]
where 
$
B_{1}=J_{+}\otimes a + J_{-}\otimes a^{\dagger}
$
and $\left\{J_{+},J_{-}\right\}$ are just generators of (spin one) 
irreducible representation of (\ref{eq:sigmas}). We note that this means 
a well--known decomposition of spin 
$\frac{1}{2}\otimes \frac{1}{2}=0\oplus 1$. 
From the decomposition we have 
\begin{eqnarray}
\label{eq:expo-two}
\mbox{e}^{-itgA_{2}}
=
T
\left(
  \begin{array}{cc}
  1 &                       \\
    & \mbox{e}^{-itgB_{1}}
  \end{array}
\right)
T^{\dagger}.
\end{eqnarray}
Therefore to calculate $\mbox{e}^{-itgA_{2}}$ we have only to do 
$\mbox{e}^{-itgB_{1}}$. 
Noting the relation 
\begin{eqnarray}
B_{1}^{2}&=&
\left(
  \begin{array}{ccc}
    2(N+1)             &    0     & 2a^{2}    \\
      0                & 2(2N+1)  &  0        \\
    2(a^{\dagger})^{2} &    0     & 2N
  \end{array}
\right),          \nonumber \\
\label{eq:relation-two}
B_{1}^{3}&=&
\left(
  \begin{array}{ccc}
    2(2N+3) &         &          \\
            & 2(2N+1) &          \\
            &         & 2(2N-1)
  \end{array}
\right)B_{1}\equiv DB_{1},  
\end{eqnarray}
and so 
\[
B_{1}^{2n}=D^{n-1}B_{1}^{2}\quad \mbox{for}\quad n\geq 1,\quad 
B_{1}^{2n+1}=D^{n}B_{1}\quad \mbox{for}\quad n\geq 0
\]
we obtain by making use of the Taylor expansion 
\begin{eqnarray}
\label{eq:exponential-1}
\mbox{e}^{-itgB_{1}}
&=&
{\bf 1}+ \sum_{n=1}^{\infty}\frac{(-1)^n}{(2n)!}
 \left(tg\right)^{2n}B_{1}^{2n}
-i\sum_{n=0}^{\infty}\frac{(-1)^n}{(2n+1)!}
 \left(tg\right)^{2n+1}B_{1}^{2n+1}        \nonumber \\
&=&
\left(
  \begin{array}{ccc}
   1+\frac{2N+2}{2N+3}f(N+1) & -ih(N+1)a & \frac{2}{2N+3}f(N+1)a^{2}  \\
   -ih(N)a^{\dagger} & 1+2f(N) & -ih(N)a                              \\
   \frac{2}{2N-1}f(N-1)(a^{\dagger})^{2} & -ih(N-1)a^{\dagger} & 
   1+\frac{2N}{2N-1}f(N-1)
  \end{array}
\right)
\end{eqnarray}
where 
\[
f(N)=\frac{-1+\mbox{cos}\left(tg\sqrt{2(2N+1)}\right)}{2},\quad 
h(N)=\frac{\mbox{sin}\left(tg\sqrt{2(2N+1)}\right)}{\sqrt{2N+1}}.
\]

\vspace{3mm}
\par \noindent 
{\bf Three Atoms Case}\quad In this case $A$ in (\ref{eq:A}) is written as 
\begin{equation}
\label{eq:A-three}
A_{3}=
\left(
  \begin{array}{cccccccc}
    0 &          a & a &           0  & a & 0 & 0 & 0          \\
    a^{\dagger}& 0 & 0 &           a  & 0 & a & 0 & 0          \\
    a^{\dagger}& 0 & 0 &           a  & 0 & 0 & a & 0          \\
    0 & a^{\dagger}& a^{\dagger} & 0  & 0 & 0 & 0 & a          \\
    a^{\dagger}& 0 & 0  &  0          & 0 & a & a & 0          \\
    0 & a^{\dagger}& 0  & 0   & a^{\dagger} &  0 & 0 & a       \\
    0 & 0 & a^{\dagger} & 0  & a^{\dagger} &  0 & 0 & a        \\
    0 & 0 & 0 & a^{\dagger} & 0 & a^{\dagger} & a^{\dagger} & 0    
  \end{array}
\right).
\end{equation}

We would like to look for the explicit form of solution like 
(\ref{eq:exponential-1/2}), (\ref{eq:exponential-1}). 
If we set 
\begin{equation}
\label{eq:T-three}
T=
\left(
  \begin{array}{cccccccc}
    0 & 0 & 0 & 0 & 1 & 0 & 0 & 0 \\
    \frac{1}{\sqrt{2}} & 0 & \frac{1}{\sqrt{6}} & 0 & 0 & 
    \frac{1}{\sqrt{3}} & 0 & 0 \\
    -\frac{1}{\sqrt{2}} & 0 & \frac{1}{\sqrt{6}} & 0 & 0 & 
    \frac{1}{\sqrt{3}} & 0 & 0 \\ 
    0 & 0 & 0 & \frac{\sqrt{2}}{\sqrt{3}} & 0 & 0 & \frac{1}{\sqrt{3}} & 0 \\
    0 & 0 & -\frac{\sqrt{2}}{\sqrt{3}} & 0 & 0 & \frac{1}{\sqrt{3}} & 0 & 0 \\
    0 & \frac{1}{\sqrt{2}} & 0 & -\frac{1}{\sqrt{6}} & 0 & 0 & 
    \frac{1}{\sqrt{3}} & 0 \\
    0 & -\frac{1}{\sqrt{2}} & 0 & -\frac{1}{\sqrt{6}} & 0 & 0 & 
    \frac{1}{\sqrt{3}} & 0 \\
    0 & 0 & 0 & 0 & 0 & 0 & 0 & 1
  \end{array}
\right),
\end{equation}
then it is not difficult to see 
\[
T^{\dagger}A_{3}T=
\left(
  \begin{array}{cccccccc}
     0 & a &   &   &   &   &   &                       \\
    a^{\dagger}& 0 &   &   &    &   &   &              \\
       &   & 0 &  a &   &   &    &                     \\
       &   & a^{\dagger} & 0 &   &   &   &             \\
       &   &   &   & 0 & \sqrt{3}a & 0 & 0             \\
       &   &   &   & \sqrt{3}a^{\dagger} & 0 & 2a & 0  \\
       &   &   &   & 0 & 2a^{\dagger} & 0 & \sqrt{3}a  \\   
       &   &   &   & 0 & 0 & \sqrt{3}a^{\dagger} & 0
  \end{array}
\right)\equiv 
\left(
  \begin{array}{ccc}
     B_{1/2} &       &        \\
           & B_{1/2} &        \\ 
           &       & B_{3/2}
  \end{array}
\right).
\]
This means a decomposition of spin $\frac{1}{2}\otimes \frac{1}{2}\otimes 
\frac{1}{2}=\frac{1}{2}\oplus \frac{1}{2}\oplus \frac{3}{2}$. 
From the decomposition we have 
\begin{equation}
\label{eq:expo-three}
\mbox{e}^{-itgA_{3}}
=
T
\left(
  \begin{array}{ccc}
     \mbox{e}^{-itgB_{1/2}} &       &        \\
           & \mbox{e}^{-itgB_{1/2}} &        \\ 
           &       & \mbox{e}^{-itgB_{3/2}}
  \end{array}
\right)
T^{\dagger}.
\end{equation}
Therefore we have only to calculate $\mbox{e}^{-itgB_{3/2}}$, which is however 
not easy. 
In this case there is no simple relation like (\ref{eq:relation-one}) or 
(\ref{eq:relation-two}), so we must find another one. 

Let us state {\bf the key lemma} for that. Noting 
\begin{eqnarray}
B_{3/2}^{2}
&=&
\left(
  \begin{array}{cccc}
    3N+3 & 0 & 2\sqrt{3}a^{2} & 0              \\
     0 & 7N+4 & 0 & 2\sqrt{3}a^{2}             \\
     2\sqrt{3}(a^{\dagger})^{2} & 0 & 7N+3 & 0 \\
     0 & 2\sqrt{3}(a^{\dagger})^{2} & 0 & 3N 
  \end{array}
\right),       \nonumber \\
B_{3/2}^{3}
&=&
\left(
  \begin{array}{cccc}
    0 & \sqrt{3}(7N+11)a & 0 & 6a^{3}                       \\
    \sqrt{3}(7N+4)a^{\dagger} & 0 & 20(N+1)a & 0            \\
    0 & 20Na^{\dagger} & 0 & \sqrt{3}(7N+3)a                \\
    6(a^{\dagger})^{3} & 0 & \sqrt{3}(7N-4)a^{\dagger} & 0
  \end{array}
\right),       \nonumber 
\end{eqnarray}
and the relations 
\[
B_{3/2}^{2n+1}=B_{3/2}B_{3/2}^{2n},\quad 
B_{3/2}^{2n+2}=B_{3/2}^{2}B_{3/2}^{2n},
\]
we can obtain $B_{3/2}^{2n}$ and $B_{3/2}^{2n+1}$ like 
\begin{eqnarray}
\label{eq:even-relation}
B_{3/2}^{2n}
&=&
\left(
  \begin{array}{cccc}
    \alpha_{n}(N+2) & 0 & 2\sqrt{3}\xi_{n}(N+2)a^{2} & 0               \\
     0 & \beta_{n}(N+1) & 0 & 2\sqrt{3}\xi_{n}(N+1)a^{2}               \\
     2\sqrt{3}\xi_{n}(N)(a^{\dagger})^{2} & 0 & \gamma_{n}(N) & 0      \\
     0 & 2\sqrt{3}\xi_{n}(N-1)(a^{\dagger})^{2} & 0 & \delta_{n}(N-1)
  \end{array}
\right),       \nonumber \\
&&  \\
\label{eq:odd-relation}
B_{3/2}^{2n+1}
&=&
\left(
  \begin{array}{cccc}
    0 & \sqrt{3}\beta_{n}(N+2)a & 0 & 6\xi_{n}(N+2)a^{3}          \\
    \sqrt{3}\beta_{n}(N+1)a^{\dagger} & 0 & 2\xi_{n+1}(N+1)a & 0  \\
    0 & 2\xi_{n+1}(N)a^{\dagger} & 0 & \sqrt{3}\gamma_{n}(N)a     \\
    6\xi_{n}(N-1)(a^{\dagger})^{3} & 0 & \sqrt{3}\gamma_{n}(N-1)a^{\dagger} & 0
  \end{array}
\right), 
\end{eqnarray}
where 
\begin{eqnarray*}
\alpha_{n}(N)&=&(v_{+}\lambda_{+}^{n}-v_{-}\lambda_{-}^{n})/(2\sqrt{d}), \quad 
\beta_{n}(N)=(w_{+}\lambda_{+}^{n}-w_{-}\lambda_{-}^{n})/(2\sqrt{d}),  \\
\gamma_{n}(N)&=&(v_{+}\lambda_{-}^{n}-v_{-}\lambda_{+}^{n})/(2\sqrt{d}), \quad 
\delta_{n}(N)=(w_{+}\lambda_{-}^{n}-w_{-}\lambda_{+}^{n})/(2\sqrt{d}), \\
\xi_{n}(N)&=&(\lambda_{+}^{n}-\lambda_{-}^{n})/(2\sqrt{d}), 
\end{eqnarray*}
and $\lambda_{\pm}\equiv \lambda_{\pm}(N)$, $v_{\pm}\equiv v_{\pm}(N)$, 
$w_{\pm}\equiv w_{\pm}(N)$, $d\equiv d(N)$ defined by 
\begin{eqnarray*}
\lambda_{\pm}(N)&=&5N\pm \sqrt{d(N)},\ 
v_{\pm}(N)=-2N-3\pm \sqrt{d(N)},\ 
w_{\pm}(N)=2N-3\pm \sqrt{d(N)},         \\
d(N)&=&16N^{2}+9.
\end{eqnarray*}

Then by making use of (\ref{eq:even-relation}) and (\ref{eq:odd-relation}) 
we have 
\begin{eqnarray}
\label{eq:exponential-3/2}
&&\mbox{e}^{-itgB_{3/2}}   
=
 \sum_{n=0}^{\infty}\frac{(-1)^n}{(2n)!}
 \left(tg\right)^{2n}B_{3/2}^{2n}
-i\sum_{n=0}^{\infty}\frac{(-1)^n}{(2n+1)!}
 \left(tg\right)^{2n+1}B_{3/2}^{2n+1}         \nonumber \\
=&&
\left(
  \begin{array}{cccc}
    f_{2}(N+2) & -\sqrt{3}iF_{1}(N+2)a & 2\sqrt{3}h_{1}(N+2)a^{2} & 
    -6iH_{0}(N+2)a^{3}   \\
    -\sqrt{3}iF_{1}(N+1)a^{\dagger} & f_{1}(N+1) & -2iH_{1}(N+1)a & 
    2\sqrt{3}h_{1}(N+1)a^{2} \\
    2\sqrt{3}h_{1}(N)(a^{\dagger})^{2} & -2iH_{1}(N)a^{\dagger} & 
    f_{0}(N) & -\sqrt{3}iF_{0}(N)a  \\
    -6iH_{0}(N-1)(a^{\dagger})^{3} & 2\sqrt{3}h_{1}(N-1)(a^{\dagger})^{2} & 
    -\sqrt{3}iF_{0}(N-1)a^{\dagger} & f_{-1}(N-1)   
  \end{array}
\right)      \nonumber \\ 
&{}& 
\end{eqnarray}
where 
\begin{eqnarray}
f_{2}(N)&=&\left\{v_{+}(N)\mbox{cos}(tg\sqrt{\lambda_{+}(N)})-
v_{-}(N)\mbox{cos}(tg\sqrt{\lambda_{-}(N)})\right\}/(2\sqrt{d(N)}), 
\nonumber \\
f_{1}(N)&=&\left\{w_{+}(N)\mbox{cos}(tg\sqrt{\lambda_{+}(N)})-
w_{-}(N)\mbox{cos}(tg\sqrt{\lambda_{-}(N)})\right\}/(2\sqrt{d(N)}), 
\nonumber \\
f_{0}(N)&=&\left\{v_{+}(N)\mbox{cos}(tg\sqrt{\lambda_{-}(N)})-
v_{-}(N)\mbox{cos}(tg\sqrt{\lambda_{+}(N)})\right\}/(2\sqrt{d(N)}), 
\nonumber \\
f_{-1}(N)&=&\left\{w_{+}(N)\mbox{cos}(tg\sqrt{\lambda_{-}(N)})-
w_{-}(N)\mbox{cos}(tg\sqrt{\lambda_{+}(N)})\right\}/(2\sqrt{d(N)}), 
\nonumber \\
h_{1}(N)&=&\left\{\mbox{cos}(tg\sqrt{\lambda_{+}(N)})-
\mbox{cos}(tg\sqrt{\lambda_{-}(N)})\right\}/(2\sqrt{d(N)}),
\nonumber \\
F_{1}(N)&=&\left\{\frac{w_{+}(N)}{\sqrt{\lambda_{+}(N)}}
\mbox{sin}(tg\sqrt{\lambda_{+}(N)})-
\frac{w_{-}(N)}{\sqrt{\lambda_{-}(N)}}
\mbox{sin}(tg\sqrt{\lambda_{-}(N)})\right\}/(2\sqrt{d(N)}), 
\nonumber \\
F_{0}(N)&=&\left\{\frac{v_{+}(N)}{\sqrt{\lambda_{-}(N)}}
\mbox{sin}(tg\sqrt{\lambda_{-}(N)})-
\frac{v_{-}(N)}{\sqrt{\lambda_{+}(N)}}
\mbox{sin}(tg\sqrt{\lambda_{+}(N)})\right\}/(2\sqrt{d(N)}), 
\nonumber \\
H_{1}(N)&=&\left\{\sqrt{\lambda_{+}(N)}
\mbox{sin}(tg\sqrt{\lambda_{+}(N)})-
\sqrt{\lambda_{-}(N)}
\mbox{sin}(tg\sqrt{\lambda_{-}(N)})\right\}/(2\sqrt{d(N)}),
\nonumber \\
H_{0}(N)&=&\left\{\frac{1}{\sqrt{\lambda_{+}(N)}}
\mbox{sin}(tg\sqrt{\lambda_{+}(N)})-
\frac{1}{\sqrt{\lambda_{-}(N)}}
\mbox{sin}(tg\sqrt{\lambda_{-}(N)})\right\}/(2\sqrt{d(N)}).
\nonumber 
\end{eqnarray}

\vspace{10mm}
Now we must calculate the term 
\begin{equation}
\label{eq:full-term}
F(t)\equiv 
\mbox{e}^{itgA_{n}}
\left(\mbox{e}^{it\omega S_{3}}\otimes \mbox{e}^{it\omega N}\right)
V(t)
\left(\mbox{e}^{-it\omega S_{3}}\otimes \mbox{e}^{-it\omega N}\right)
\mbox{e}^{-itgA_{n}}
\end{equation}
from (\ref{eq:reduction-equation}), so we introduce a brief notation 
\[
\tilde{V}(t)\equiv 
\left(\mbox{e}^{it\omega S_{3}}\otimes \mbox{e}^{it\omega N}\right)
V(t)
\left(\mbox{e}^{-it\omega S_{3}}\otimes \mbox{e}^{-it\omega N}\right).
\]
Therefore we want to calculate 
\[
F(t)
= \mbox{e}^{itgA_{n}}\tilde{V}(t)\mbox{e}^{-itgA_{n}}
=T(\mbox{block diagonals})T^{\dagger}\tilde{V}(t)T
 (\mbox{block diagonals})^{\dagger}T^{\dagger}
\]
explicitly for the case of $n=2$ and $3$. For that let us calculate 
$\tilde{V}(t)$ and $T^{\dagger}\tilde{V}(t)T$ in advance. 
The calculation is straightforward and the result is 

\vspace{3mm}
\par \noindent 
{\bf Two Atoms Case}\quad 

\begin{eqnarray}
\label{eq:S-N-V-S-N-two}
&&
\tilde{V}(t)=
\left(
  \begin{array}{cccc}
    0 & q(t) & p(t) & 0               \\
    \bar{q}(t) & 0 & 0 & p(t)         \\
    \bar{p}(t) & 0 & 0 & q(t)         \\
    0 & \bar{p}(t) & \bar{q}(t) & 0 
  \end{array}
\right)\otimes {\bf 1}  \\
&&\mbox{with}\quad 
p(t)\equiv h_{1}\mbox{e}^{i\{(\Omega_{1}+\omega)t+\phi_{1}\}},\quad 
q(t)\equiv h_{2}\mbox{e}^{i\{(\Omega_{2}+\omega)t+\phi_{2}\}}
\nonumber 
\end{eqnarray}
and 
\begin{eqnarray}
\label{eq:S-N-V-S-N-two-more}
&&
T^{\dagger}\tilde{V}(t)T=
\left(
  \begin{array}{cccc}
  0 & \frac{-\bar{p}+\bar{q}}{\sqrt{2}} & 0 & \frac{p-q}{\sqrt{2}}  \\
  \frac{-p+q}{\sqrt{2}} & 0 & \frac{p+q}{\sqrt{2}} & 0              \\
  0 & \frac{\bar{p}+\bar{q}}{\sqrt{2}} & 0 & \frac{p+q}{\sqrt{2}}   \\
  \frac{\bar{p}-\bar{q}}{\sqrt{2}} & 0 & \frac{\bar{p}+\bar{q}}{\sqrt{2}} & 0 
  \end{array}
\right)\otimes {\bf 1} 
\end{eqnarray}
where we have omitted the time $t$ for simplicity.

\vspace{3mm}
\par \noindent 
{\bf Three Atoms Case}\quad 

\begin{eqnarray}
\label{eq:S-N-V-S-N-three}
&&
\tilde{V}(t)=
\left(
  \begin{array}{cccccccc}
  0 & r(t) & q(t) & 0 & p(t) & 0 & 0 & 0                   \\
  \bar{r}(t) & 0 & 0 & q(t) & 0 & p(t) & 0 & 0             \\
  \bar{q}(t) & 0 & 0 & r(t) & 0 & 0 & p(t) & 0             \\
  0 & \bar{q}(t) & \bar{r}(t) & 0 & 0 & 0 & 0 & p(t)       \\
  \bar{p}(t) & 0 & 0 & 0 & 0 & r(t) & q(t) & 0             \\
  0 & \bar{p}(t) & 0 & 0 & \bar{r}(t) & 0 & 0 & q(t)       \\
  0 & 0 & \bar{p}(t) & 0 & \bar{q}(t) & 0 & 0 & r(t)       \\
  0 & 0 & 0 & \bar{p}(t) & 0 & \bar{q}(t) & \bar{r}(t) & 0
  \end{array}
\right)\otimes {\bf 1} \\
&&\mbox{with}\quad 
p(t)\equiv h_{1}\mbox{e}^{i\{(\Omega_{1}+\omega)t+\phi_{1}\}},\quad 
q(t)\equiv h_{2}\mbox{e}^{i\{(\Omega_{2}+\omega)t+\phi_{2}\}},\quad
r(t)\equiv h_{3}\mbox{e}^{i\{(\Omega_{3}+\omega)t+\phi_{3}\}}
\nonumber 
\end{eqnarray}
and
\begin{eqnarray}
\label{eq:S-N-V-S-N-three-more}
&&
T^{\dagger}\tilde{V}(t)T=
\left(
  \begin{array}{cccccccc}
  0 & p & 0 & \frac{q-r}{\sqrt{3}} & \frac{-\bar{q}+\bar{r}}{\sqrt{2}} & 0 & 
  \frac{q-r}{\sqrt{6}} & 0       \\
  \bar{p} & 0 & \frac{\bar{q}-\bar{r}}{\sqrt{3}} & 0 & 0 & 
  \frac{-\bar{q}+\bar{r}}{\sqrt{6}} & 0 & \frac{q-r}{\sqrt{2}}   \\
  0 & \frac{q-r}{\sqrt{3}} & 0 & \frac{-p+2q+2r}{3} & 
  \frac{-2\bar{p}+\bar{q}+\bar{r}}{\sqrt{6}} & 0 & \frac{2p-q-r}{3\sqrt{2}} & 
  0  \\
  \frac{\bar{q}-\bar{r}}{\sqrt{3}}& 0 & \frac{-\bar{p}+2\bar{q}+2\bar{r}}{3}
  & 0 & 0 & \frac{-2\bar{p}+\bar{q}+\bar{r}}{3\sqrt{2}} & 0 & \frac{2p-q-r}
  {\sqrt{6}}  \\
  \frac{-q+r}{\sqrt{2}} & 0 & \frac{-2p+q+r}{\sqrt{6}} & 0 & 0 & 
  \frac{p+q+r}{\sqrt{3}} & 0 & 0             \\
  0 & \frac{-q+r}{\sqrt{6}} & 0 & \frac{-2p+q+r}{3\sqrt{2}} & 
  \frac{\bar{p}+\bar{q}+\bar{r}}{\sqrt{3}} & 0 & \frac{2(p+q+r)}{3} & 0  \\
  \frac{\bar{q}-\bar{r}}{\sqrt{6}} & 0 & \frac{2\bar{p}-\bar{q}-\bar{r}}
  {3\sqrt{2}} & 0 & 0 & \frac{2(\bar{p}+\bar{q}+\bar{r})}{3} & 0 & 
  \frac{p+q+r}{\sqrt{3}}      \\
  0 & \frac{\bar{q}-\bar{r}}{\sqrt{2}} & 0 & \frac{2\bar{p}-\bar{q}-\bar{r}}
  {\sqrt{6}} & 0 & 0 & \frac{\bar{p}+\bar{q}+\bar{r}}{\sqrt{3}} & 0
  \end{array}
\right)\otimes {\bf 1}   \nonumber \\
&&{}
\end{eqnarray}
where we have omitted the time $t$ for simplicity.

\vspace{5mm}
To write down all components of 
$F(t)=\mbox{e}^{itgA_{n}}\tilde{V}(t)\mbox{e}^{-itgA_{n}}$ is very long and 
we moreover don't need all of them, so we omit it here. 
Next let us go to our quantum computation based on a few atoms of 
laser--cooled and trapped linearly in a cavity.

\section{Quantum Computation} 

To develop a quantum computation based on atoms laser--cooled and trapped 
{\bf linearly} in a cavity (Cavity QED Quantum Computation) 
a quick and clear construction of the both controlled NOT gate and 
controlled--controlled NOT gate is required, \cite{nine}. 
Let us construct such very important quantum gates in this section.

\subsection{Controlled NOT Gate}

In this subsection we treat the case of two atoms (the system of two qubits). 
First let us make a short review of the system of two--qubits. Each element 
can be written as 
\[
\psi=a_{++}\ket{+}\otimes \ket{+}+a_{+-}\ket{+}\otimes \ket{-}+
a_{-+}\ket{-}\otimes \ket{+}+a_{--}\ket{-}\otimes \ket{-}
\]
with two bases $\ket{+}$ and $\ket{-}$ and $|a_{++}|^{2}+|a_{+-}|^{2}+
|a_{-+}|^{2}+|a_{--}|^{2}=1$. Here if we identify 
\[
\ket{+}=
\left(
  \begin{array}{c}
   1 \\
   0
  \end{array}
\right),\quad 
\ket{-}=
\left(
  \begin{array}{c}
   0 \\
   1
  \end{array}
\right),
\]
then $\psi$ above becomes 
\begin{equation}
\label{eq:}
\psi=
\left(
  \begin{array}{c}
    a_{++} \\
    a_{+-} \\
    a_{-+} \\
    a_{--}
  \end{array}
\right).
\end{equation}

How do we embed two--qubits in our quantized system ? It is not known at the 
current time, which may depend on methods of experimentalists. 
Therefore let us consider the simplest one like 
\begin{equation}
\label{eq:encode two qubits}
\ket{\psi(t)}=
\left(
  \begin{array}{c}
    a_{++}(t) \\
    a_{+-}(t) \\
    a_{-+}(t) \\
    a_{--}(t)
  \end{array}
\right)\otimes \ket{0}, 
\end{equation}
where $\ket{0}$ is the ground state of the radiation field ($a\ket{0}=0$). 
We note that in full theory we must consider the following superpositions 
\[
\ket{\Psi(t)}=\sum_{n=0}^{\infty}
\left(
  \begin{array}{c}
    a_{++,n}(t) \\
    a_{+-,n}(t) \\
    a_{-+,n}(t) \\
    a_{--,n}(t)
  \end{array}
\right)\otimes \ket{n} 
\]
as a wave function, which is however too complicated to solve. 

To determine a dynamics that the coefficients 
$a_{++},a_{+-},a_{-+},a_{--}$ will satisfy we substitute 
(\ref{eq:encode two qubits}) into the equation 
\begin{eqnarray*}
i\frac{d}{dt}\ket{\psi(t)}
&=&F(t)\ket{\psi(t)}
=\mbox{e}^{itgA_{2}}\tilde{V}(t)\mbox{e}^{-itgA_{2}}\ket{\psi(t)}
    \nonumber \\
&=&
T
\left(
  \begin{array}{cc}
  1 &                       \\
    & \mbox{e}^{itgB_{1}}
  \end{array}
\right)
T^{\dagger}\tilde{V}(t)T
\left(
  \begin{array}{cc}
  1 &                       \\
    & \mbox{e}^{-itgB_{1}}
  \end{array}
\right)
T^{\dagger}\ket{\psi(t)}.
\end{eqnarray*}
Let us rewrite the above equation. If we set 
\begin{equation}
\label{eq:ansatz-modified-two}
\ket{\varphi(t)}\equiv T^{\dagger}\ket{\psi(t)} \Longleftrightarrow 
\left(
  \begin{array}{c}
    \varphi_{0}(t) \\
    \varphi_{1}(t) \\
    \varphi_{2}(t) \\
    \varphi_{3}(t) 
  \end{array}
\right)\otimes \ket{0}
\equiv
\left(
  \begin{array}{c}
    \frac{a_{+-}(t)-a_{-+}(t)}{\sqrt{2}}  \\
    a_{++}(t)                             \\
    \frac{a_{+-}(t)+a_{-+}(t)}{\sqrt{2}}  \\
    a_{--}(t)
  \end{array}
\right)\otimes \ket{0}
\end{equation}
then 
\begin{equation}
\label{eq:reduction-equation-two}
i\frac{d}{dt}\ket{\varphi(t)}=
\left(
  \begin{array}{cc}
  1 &                       \\
    & \mbox{e}^{itgB_{1}}
  \end{array}
\right)
T^{\dagger}\tilde{V}(t)T
\left(
  \begin{array}{cc}
  1 &                       \\
    & \mbox{e}^{-itgB_{1}}
  \end{array}
\right)
\ket{\varphi(t)}.
\end{equation}
On the other hand, we have calculated the term $T^{\dagger}\tilde{V}(t)T$ 
in (\ref{eq:S-N-V-S-N-two-more}). 

Note that the above equation is not satisfied under the restrictive ansatz 
(\ref{eq:ansatz-modified-two}). 
Because the left hand side contains only the ground state $\ket{0}$, 
while the right hand side contains the ground state $\ket{0}$ and some 
excited states $\ket{1}$, $\ket{2}$, $\ket{3}$. 
However, the states $\ket{1}$, $\ket{2}$, $\ket{3}$ which have no 
corresponding kinetic terms contain the coupling constants $h_{1}$ and 
$h_{2}$ (see $p(t)$ and $q(t)$ in (\ref{eq:S-N-V-S-N-two})), 
so the equation is approximately satisfied if they are {\bf small enough} 
(namely, in the weak coupling regime in the title).

\par \noindent
Therefore the (full) equation is reduced to the equations of 
$\{\varphi_{0},\varphi_{1},\varphi_{2},\varphi_{3}\}$ at the ground state 
$\ket{0}$ : 
\begin{equation}
\label{eq:matrix-form}
i\frac{d}{dt}
\left(
  \begin{array}{c}
    \varphi_{0}(t) \\
    \varphi_{1}(t) \\
    \varphi_{2}(t) \\
    \varphi_{3}(t)
  \end{array}
\right)
=
\left(
  \begin{array}{cccc}
    0     & x_{12} &  0     & x_{14}  \\
   x_{21} &   0    & x_{23} & 0       \\
    0     & x_{32} &  0     & x_{34}  \\
   x_{41} & 0      & x_{43} & 0
  \end{array}
\right)
\left(
  \begin{array}{c}
    \varphi_{0}(t) \\
    \varphi_{1}(t) \\
    \varphi_{2}(t) \\
    \varphi_{3}(t)
  \end{array}
\right)
\end{equation}
where 
\begin{eqnarray*}
x_{12}&=&\bar{x}_{21},\\
x_{14}&=&\frac{p-q}{\sqrt{2}},\\
x_{21}&=&\frac{-p+q}{\sqrt{2}}\left(1+\frac{2}{3}f(1)\right),\\
x_{23}&=&\frac{p+q}{\sqrt{2}}\left(1+2f(0)+\frac{2}{3}f(1)+\frac{4}{3}f(0)f(1)+
h(0)h(1)\right),\\
x_{32}&=&\bar{x}_{23},\\
x_{34}&=&\frac{p+q}{\sqrt{2}}\left(1+2f(0)\right),\\
x_{41}&=&\bar{x}_{14},\\
x_{43}&=&\bar{x}_{34}
\end{eqnarray*}
and
\begin{eqnarray*}
p&=&h_{1}\mbox{e}^{i\{(\Omega_{1}+\omega)t+\phi_{1}\}},\quad
q=h_{2}\mbox{e}^{i\{(\Omega_{2}+\omega)t+\phi_{2}\}}  \\
f(0)&=&\frac{-1+\mbox{cos}(\sqrt{2}gt)}{2},\quad
f(1)=\frac{-1+\mbox{cos}(\sqrt{6}gt)}{2},\\
h(0)&=&\mbox{sin}(\sqrt{2}gt),\quad
h(1)=\frac{\mbox{sin}(\sqrt{6}gt)}{\sqrt{3}}.
\end{eqnarray*}

How do we solve it ? We use some resonance condition and the rotating wave 
approximation associated to it, which is popular in quantum optics or in a 
field of laser physics. Let us focus on the (2,3)--component of the matrix 
which came from the interaction term. 
The products $f(0)f(1)$ and $h(0)h(1)$ contain the term 
$
\mbox{e}^{-itg(\sqrt{2}+\sqrt{6})}
$
by the Euler formulas 
$
\mbox{cos}(\theta)=(\mbox{e}^{i\theta}+\mbox{e}^{-i\theta})/2,\ 
\mbox{sin}(\theta)=(\mbox{e}^{i\theta}-\mbox{e}^{-i\theta})/2i
$. 
Noting 
\[
\mbox{e}^{i\{(\Omega_{1}+\omega)t+\phi_{1}\}}
\mbox{e}^{-itg(\sqrt{2}+\sqrt{6})}
=
\mbox{e}^{i\{(\Omega_{1}+\omega-(\sqrt{2}+\sqrt{6})g)t+\phi_{1}\}}, 
\]
we set a new resonance condition 
\begin{equation}
\label{eq:new-resonance}
\Omega_{1}+\omega-(\sqrt{2}+\sqrt{6})g=0.
\end{equation}
All terms in (\ref{eq:matrix-form}) except for the constant one 
$
\mbox{e}^{i\{(\Omega_{1}+\omega-(\sqrt{2}+\sqrt{6})g)t+\phi_{1}\}}=
\mbox{e}^{i\phi_{1}}
$ 
contain ones like $\mbox{e}^{i(t\theta+\alpha)}$ ($\theta\neq 0$), 
so we neglect all such oscillating terms (a rotating wave approximation). 
Then (\ref{eq:matrix-form}) reduces to a very simple matrix equation 
\begin{equation}
\label{eq:reduced-matrix-form}
i\frac{d}{dt}
\left(
  \begin{array}{c}
    \varphi_{0}(t) \\
    \varphi_{1}(t) \\
    \varphi_{2}(t) \\
    \varphi_{3}(t)
  \end{array}
\right)
=
\frac{-\sqrt{2}(\sqrt{3}-1)h_{1}}{24}
\left(
  \begin{array}{cccc}
      0 & 0                     & 0                    & 0  \\
      0 & 0                     & \mbox{e}^{i\phi_{1}} & 0  \\
      0 & \mbox{e}^{-i\phi_{1}} & 0                    & 0  \\
      0 & 0                     & 0                    & 0
  \end{array}
\right)
\left(
  \begin{array}{c}
    \varphi_{0}(t) \\
    \varphi_{1}(t) \\
    \varphi_{2}(t) \\
    \varphi_{3}(t)
  \end{array}
\right).
\end{equation}
The solution is easily obtained to be 
\begin{eqnarray}
\label{eq:solution}
\left(
  \begin{array}{c}
    \varphi_{0}(t) \\
    \varphi_{1}(t) \\
    \varphi_{2}(t) \\
    \varphi_{3}(t)
  \end{array}
\right)
&=&
\mbox{exp}
\left\{
\frac{i(\sqrt{6}-\sqrt{2})h_{1}t}{24}
\left(
  \begin{array}{cccc}
      0 & 0                     & 0                    & 0  \\
      0 & 0                     & \mbox{e}^{i\phi_{1}} & 0  \\
      0 & \mbox{e}^{-i\phi_{1}} & 0                    & 0  \\
      0 & 0                     & 0                    & 0
  \end{array}
\right)
\right\}
\left(
  \begin{array}{c}
    \varphi_{0}(0) \\
    \varphi_{1}(0) \\
    \varphi_{2}(0) \\
    \varphi_{3}(0)
  \end{array}
\right)           \nonumber \\
&=&
\left(
  \begin{array}{cccc}
  1 & 0 & 0 & 0                                                            \\
  0 & \mbox{cos}(\alpha t) & i\mbox{e}^{i\phi_{1}}\mbox{sin}(\alpha t) & 0 \\ 
  0 & i\mbox{e}^{-i\phi_{1}}\mbox{sin}(\alpha t) & \mbox{cos}(\alpha t) & 0 \\
  0 & 0 & 0 & 1
  \end{array}
\right)
\left(
  \begin{array}{c}
    \varphi_{0}(0) \\
    \varphi_{1}(0) \\
    \varphi_{2}(0) \\
    \varphi_{3}(0)
  \end{array}
\right)           \nonumber \\
&\equiv&
U(t)
\left(
  \begin{array}{c}
    \varphi_{0}(0) \\
    \varphi_{1}(0) \\
    \varphi_{2}(0) \\
    \varphi_{3}(0)
  \end{array}
\right)
\end{eqnarray}
where we have set $\alpha=\frac{\sqrt{6}-\sqrt{2}}{24}h_{1}$.
That is, we obtained the unitary operator $U(t)$. In particular, if we 
choose $t_{0}$ and $\phi_{1}$ satisfying 
\[
\mbox{cos}(\alpha t_{0})=0\ (\mbox{sin}(\alpha t_{0})=1) \quad 
\mbox{and}\quad \mbox{e}^{i\phi_{1}}=i
\]
then 
\begin{equation}
U(t_{0})=
\left(
  \begin{array}{cccc}
     1 &     &      &      \\
       &  0  &  -1  &      \\
       &  1  &   0  &      \\
       &     &      &  1
  \end{array}
\right).
\end{equation}
From this we want to construct the controlled NOT operator. However, it is not 
easy \footnote{$U(t_{0})$ is imprimitive in the sense of \cite{BB}, so the 
main theorem in it says that our system is universal (namely, we can construct 
any element in $U(4)$). However, how to construct a unitary element explicitly 
is not given in \cite{BB}}.

\vspace{3mm}
At this stage we use a very skillful method due to Dirac \cite{Dirac}
. That is, we exchange two atoms in the cavity 

\vspace{5mm}
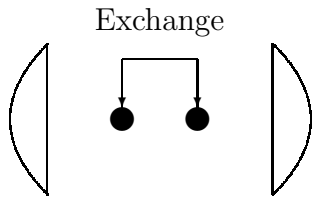
\begin{center}
\setlength{\unitlength}{1mm} 
\begin{picture}(70,40)(0,-20)
\bezier{200}(20,0)(10,10)(20,20)
\put(20,0){\line(0,1){20}}
\put(30,18){\vector(0,-1){6.5}}
\put(30,10){\circle*{3}}
\put(40,18){\vector(0,-1){6.5}}
\put(40,10){\circle*{3}}
\put(25,18){\makebox(20,10)[c]{Exchange}}
\put(30,18){\line(1,0){10}}
\bezier{200}(50,0)(60,10)(50,20)
\put(50,0){\line(0,1){20}}
\end{picture}
\end{center}
%
\vspace{-20mm}
\par \noindent
which introduces the exchange (swap) operator 
\begin{equation}
P=
\left(
  \begin{array}{cccc}
     1 &     &      &     \\
       &  0  &   1  &     \\
       &  1  &   0  &     \\
       &     &      &  1
  \end{array}
\right).
\end{equation}
Multiplying $U(t_{0})$ by $P$ gives 
\[
PU(t_{0})=
\left(
  \begin{array}{cccc}
     1 &     &     &     \\
       &  1  &     &     \\
       &     & -1  &     \\
       &     &     &  1
  \end{array}
\right)
\]
and from this we obtain 
\begin{equation}
({\bf 1}_{2}\otimes \sigma_{1})PU(t_{0})({\bf 1}_{2}\otimes \sigma_{1})
=
\left(
  \begin{array}{cccc}
     1 &     &     &     \\
       &  1  &     &     \\
       &     &  1  &     \\
       &     &     &  -1
  \end{array}
\right).
\end{equation}
This is just the controlled $\sigma_{z}$ operator.
From this it is easy to construct the controlled NOT operator, namely 
\[
C_{NOT}=({\bf 1}_{2}\otimes W)C_{\sigma_{z}}({\bf 1}_{2}\otimes W)
=
\left(
  \begin{array}{cccc}
     1 &     &     &     \\
       &  1  &     &     \\
       &     &  0  &  1  \\
       &     &  1  &  0
  \end{array}
\right)
\]
where $W$ is the Walsh--Hadamard operator given by 
\begin{equation}
W=\frac{1}{\sqrt{2}}
\left(
  \begin{array}{cc}
    1& 1 \\
    1& -1
  \end{array}
\right)
=W^{-1}.
\end{equation}
See for example \cite{KF1}. As to a construction of $W$ by making use of Rabi 
oscillations see Appendix or \cite{KF2}. 

\par \noindent
Therefore our system is universal \cite{nine}, \cite{BB}. 

{\bf A comment is in order}.

\par \noindent
(a)\ If we choose $t_{0}$ in $U(t)$ satisfying 
\[
\mbox{cos}(\alpha t_{0})=-1\quad (\mbox{sin}(\alpha t_{0})=0)
\]
then we obtain the matrix 
\[
U(t_{0})=
\left(
  \begin{array}{cccc}
     1 &     &     &     \\
       & -1  &     &     \\
       &     & -1  &     \\
       &     &     &  1
  \end{array}
\right)
=
\left(
  \begin{array}{cc}
     1 &     \\
       & -1 
  \end{array}
\right)
\otimes 
\left(
  \begin{array}{cc}
     1 &     \\
       & -1 
  \end{array}
\right)
=
\sigma_{3}\otimes \sigma_{3}.
\]

\par \vspace{3mm} \noindent
(b)\ In place of the ansatz (\ref{eq:encode two qubits}) we can set 
for example 
\[
\ket{\psi(t)}=
\left(
  \begin{array}{c}
    a_{++}(t) \\
    a_{+-}(t) \\
       0      \\
       0
  \end{array}
\right)\otimes \ket{0}
+
\left(
  \begin{array}{c}
       0      \\
       0      \\
    a_{-+}(t) \\
    a_{--}(t)
  \end{array}
\right)\otimes \ket{1}. 
\]
Then we can trace the same line shown in this section and obtain a unitary 
operator under some resonance condition like (\ref{eq:new-resonance}). 
This is a good exercise, so we leave it to the readers.

\subsection{Controlled-Controlled NOT Gate}

In this subsection we treat the case of three atoms (the system of three 
qubits). 
To perform a quantum computation we need a (rapid) construction of the 
the controlled--controlled NOT gate. 
To construct it we need a construction of the controlled NOT gates of 
three types like 

\vspace{10mm}
%
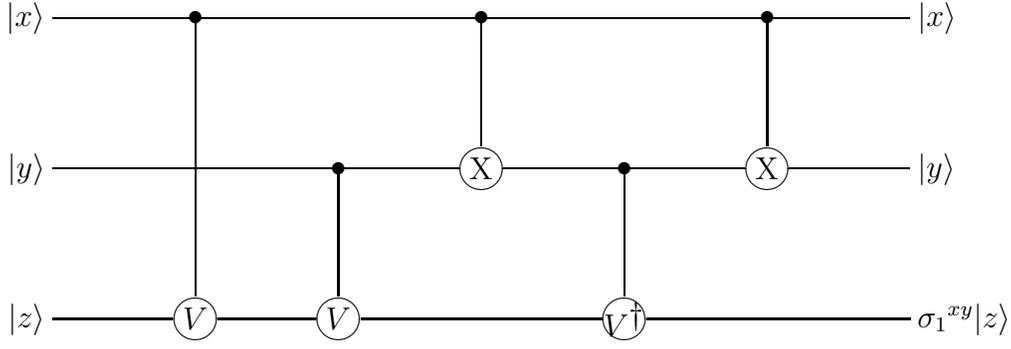
\begin{center}
\setlength{\unitlength}{1mm} 
\begin{picture}(90,40)(0,-20)
\bezier{200}(20,0)(10,10)(20,20)
\put(20,0){\line(0,1){20}}
\put(30,18){\vector(0,-1){6.5}}
\put(30,10){\circle*{3}}
\put(40,18){\vector(0,-1){6.5}}
\put(40,10){\circle*{3}}
\put(50,10){\circle*{3}}
\put(25,18){\makebox(20,10)[c]{C--NOT}}
\put(30,18){\line(1,0){10}}
\bezier{200}(60,0)(70,10)(60,20)
\put(60,0){\line(0,1){20}}
\end{picture}
\end{center}

\vspace{-15mm}
\begin{center}
\setlength{\unitlength}{1mm} 
\begin{picture}(150,40)(0,-20)
\bezier{200}(20,0)(10,10)(20,20)
\put(20,0){\line(0,1){20}}
\put(30,18){\vector(0,-1){6.5}}
\put(30,10){\circle*{3}}
\put(40,10){\circle*{3}}
\put(50,18){\vector(0,-1){6.5}}
\put(50,10){\circle*{3}}
\put(30,18){\makebox(20,10)[c]{C--NOT}}
\put(30,18){\line(1,0){20}}
\bezier{200}(60,0)(70,10)(60,20)
\put(60,0){\line(0,1){20}}
\bezier{200}(90,0)(80,10)(90,20)
\put(90,0){\line(0,1){20}}
\put(100,10){\circle*{3}}
\put(110,18){\vector(0,-1){6.5}}
\put(110,10){\circle*{3}}
\put(120,18){\vector(0,-1){6.5}}
\put(120,10){\circle*{3}}
\put(95,18){\makebox(40,10)[c]{C--NOT}}
\put(120,18){\line(-1,0){10}}
\bezier{200}(130,0)(140,10)(130,20)
\put(130,0){\line(0,1){20}}
\end{picture}
\end{center}
\vspace{-15mm}

The controlled--controlled NOT gate (operator)
\[
CC_{NOT}\ :\ \fukuso^{2} \otimes \fukuso^{2} \otimes \fukuso^{2} 
\longrightarrow \fukuso^{2} \otimes \fukuso^{2} \otimes \fukuso^{2}
\]
is shown as a picture 

\begin{center}
\setlength{\unitlength}{1mm}  
\begin{picture}(70,50)
\put(10,40){\line(1,0){30}}   
\put(10,25){\line(1,0){30}} 
\put(10,10){\line(1,0){12}} 
\put(28,10){\line(1,0){12}}   
\put(22,35){\makebox(6,10){$\bullet$}}
\put(22,20){\makebox(6,10){$\bullet$}}
\put(25,10){\circle{6}}  
\put(22,5){\makebox(6,10){$X$}} 
\put(25,40){\line(0,-1){15}}  
\put(25,25){\line(0,-1){12}}  
\put(47,20){\makebox(20,10){$=\quad {CC}_{NOT}$}}
\end{picture}
\end{center}
\vspace{-3mm}
or in a matrix form 

\[
\left(
  \begin{array}{cccccccc}
    1 &   &    &    &    &     &    &     \\
      & 1 &    &    &    &     &    &     \\
      &   & 1  &    &    &     &    &     \\
      &   &    & 1  &    &     &    &     \\
      &   &    &    & 1  &     &    &     \\
      &   &    &    &    & 1   &    &     \\
      &   &    &    &    &     &  0 & 1   \\
      &   &    &    &    &     &  1 & 0       
  \end{array}
\right).
\]
The (usual) construction by making use of controlled NOT or controlled U gates 
is shown as a picture (\cite{KF1}, \cite{nine})
%
\begin{center}
\setlength{\unitlength}{1mm}  
\begin{picture}(140,60)
\put(10,50){\line(1,0){114}} 
\put(10,30){\line(1,0){54}}  
\put(70,30){\line(1,0){32}}  
\put(108,30){\line(1,0){16}} 
\put(10,10){\line(1,0){16}}  
\put(32,10){\line(1,0){13}}  
\put(51,10){\line(1,0){32}}  
\put(89,10){\line(1,0){35}}  
\put(0,45){\makebox(9,10)[r]{$|x\rangle$}} 
\put(0,25){\makebox(9,10)[r]{$|y\rangle$}} 
\put(0,5){\makebox(9,10)[r]{$|z\rangle$}} 
\put(125,45){\makebox(25,10)[l]{$|x\rangle$}} 
\put(125,25){\makebox(25,10)[l]{$|y\rangle$}} 
\put(125,5){\makebox(25,10)[l]{${\sigma_{1}}^{xy}|z\rangle$}} 
\put(29,13){\line(0,1){37}} 
\put(48,13){\line(0,1){17}} 
\put(67,33){\line(0,1){17}} 
\put(86,13){\line(0,1){17}} 
\put(105,33){\line(0,1){17}} 
\put(26,45){\makebox(6,10){$\bullet$}} 
\put(45,25){\makebox(6,10){$\bullet$}} 
\put(64,45){\makebox(6,10){$\bullet$}} 
\put(83,25){\makebox(6,10){$\bullet$}} 
\put(102,45){\makebox(6,10){$\bullet$}} 
\put(29,10){\circle{6}}  
\put(48,10){\circle{6}}  
\put(67,30){\circle{6}}  
\put(86,10){\circle{6}}  
\put(105,30){\circle{6}} 
\put(64,25){\makebox(6,10){X}}
\put(102,25){\makebox(6,10){X}} 
\put(26,5){\makebox(6,10){$V$}} 
\put(45,5){\makebox(6,10){$V$}} 
\put(83,5){\makebox(6,10){$V^{\mbox{\dag}}$}} 
\end{picture}
\end{center}
\vspace{-5mm}
%
where $V$ is a unitary matrix given by 
\[
V=\frac{1}{2}
\left(
  \begin{array}{cc}
    1+i& 1-i \\
    1-i& 1+i
  \end{array}
\right)
\quad \Longrightarrow \quad 
V^{2}=
\left(
  \begin{array}{cc}
    0 & 1 \\
    1 & 0 
  \end{array}
\right)
=\sigma_{1}.
\]

\vspace{5mm}
However, we have not seen ``realistic" constructions in any references, so 
we give the explicit construction. See the figure 2. 

\vspace{10mm}
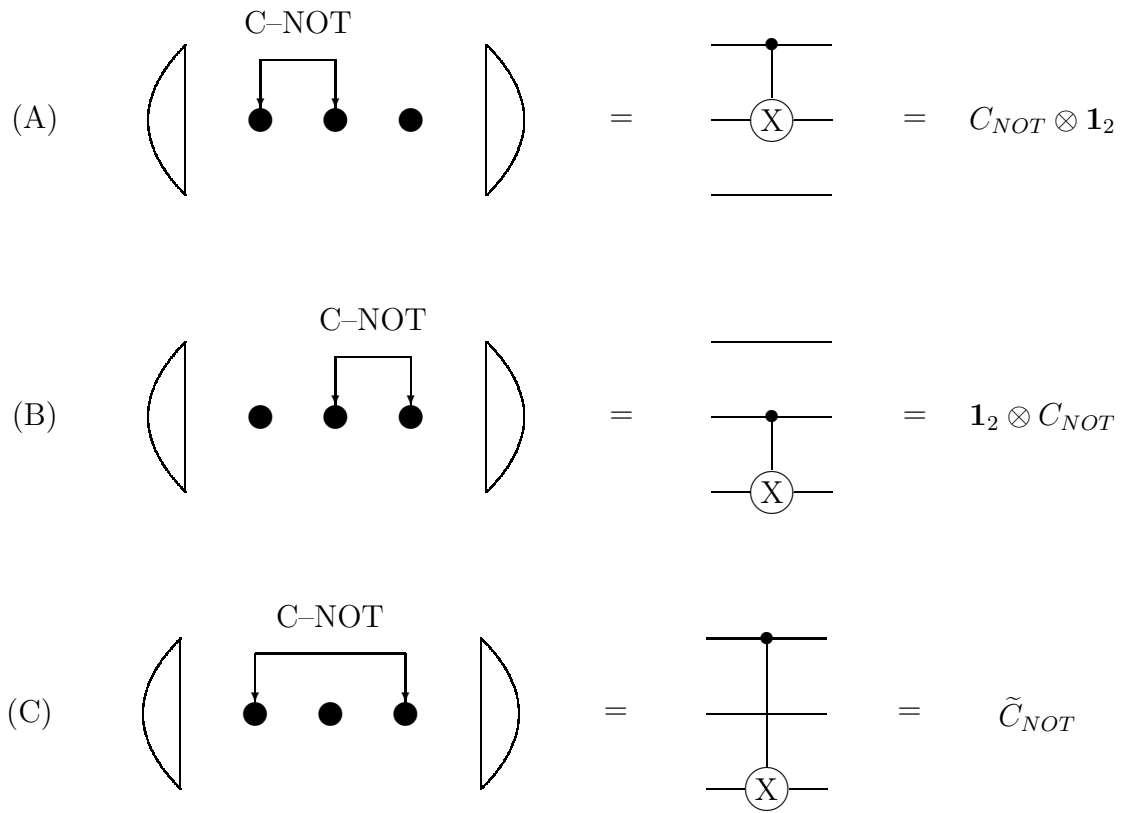
\begin{figure}
\begin{center}
\setlength{\unitlength}{1mm} 
\begin{picture}(150,40)(0,-20)
\bezier{200}(20,0)(10,10)(20,20)
\put(20,0){\line(0,1){20}}
\put(30,18){\vector(0,-1){6.5}}
\put(30,10){\circle*{3}}
\put(40,18){\vector(0,-1){6.5}}
\put(40,10){\circle*{3}}
\put(50,10){\circle*{3}}
\put(25,18){\makebox(20,10)[c]{C--NOT}}
\put(30,18){\line(1,0){10}}
\bezier{200}(60,0)(70,10)(60,20)
\put(60,0){\line(0,1){20}}
\put(0, 5){\makebox(0,10){(A)}}
\put(90,20){\line(1,0){16}}
\put(90, 0){\line(1,0){16}}
\put(90,10){\line(1,0){5}}
\put(101,10){\line(1,0){5}}
\put(95,17){\makebox(6,6){$\bullet$}} 
\put(98,10){\circle{6}}  
\put(95,7){\makebox(6,6){X}}
\put(98, 20){\line(0,-1){7}} 
\put(75,7){\makebox(6,6){$=$}} 
\put(114,7){\makebox(6,6){$=$}} 
\put(124,7){\makebox(20,6){$C_{NOT}\otimes {\bf 1}_{2}$}} 
\end{picture}
\end{center}
\vspace{-10mm}
\begin{center}
\setlength{\unitlength}{1mm} 
\begin{picture}(150,40)(0,-20)
\bezier{200}(20,0)(10,10)(20,20)
\put(20,0){\line(0,1){20}}
\put(40,18){\vector(0,-1){6.5}}
\put(30,10){\circle*{3}}
\put(50,18){\vector(0,-1){6.5}}
\put(40,10){\circle*{3}}
\put(50,10){\circle*{3}}
\put(35,18){\makebox(20,10)[c]{C--NOT}}
\put(40,18){\line(1,0){10}}
\bezier{200}(60,0)(70,10)(60,20)
\put(60,0){\line(0,1){20}}
\put(0, 5){\makebox(0,10){(B)}}
\put(90,20){\line(1,0){16}}
\put(90,10){\line(1,0){16}}
\put(90, 0){\line(1,0){5}}
\put(101,0){\line(1,0){5}}
\put(95, 7){\makebox(6,6){$\bullet$}} 
\put(98, 0){\circle{6}}  
\put(95,-3){\makebox(6,6){X}}
\put(98, 10){\line(0,-1){7}} 
\put(75,7){\makebox(6,6){$=$}} 
\put(114,7){\makebox(6,6){$=$}} 
\put(124,7){\makebox(20,6){${\bf 1}_{2}\otimes C_{NOT}$}} 
\end{picture}
\end{center}
\vspace{-10mm}
\begin{center}
\setlength{\unitlength}{1mm} 
\begin{picture}(150,40)(0,-20)
\bezier{200}(20,0)(10,10)(20,20)
\put(20,0){\line(0,1){20}}
\put(30,18){\vector(0,-1){6.5}}
\put(30,10){\circle*{3}}
\put(50,18){\vector(0,-1){6.5}}
\put(40,10){\circle*{3}}
\put(50,10){\circle*{3}}
\put(30,18){\makebox(20,10)[c]{C--NOT}}
\put(30,18){\line(1,0){20}}
\bezier{200}(60,0)(70,10)(60,20)
\put(60,0){\line(0,1){20}}
\put(0, 5){\makebox(0,10){(C)}}
\put(90,20){\line(1,0){16}}
\put(90,10){\line(1,0){16}}
\put(90, 0){\line(1,0){5}}
\put(101, 0){\line(1,0){5}}
\put(95,17){\makebox(6,6){$\bullet$}}
\put(98, 0){\circle{6}}  
\put(95,-3){\makebox(6,6){X}}
\put(98,20){\line(0,-1){17}} 
\put(75,7){\makebox(6,6){$=$}} 
\put(114,7){\makebox(6,6){$=$}} 
\put(124,7){\makebox(20,6){$\widetilde{C}_{NOT}$}} 
\end{picture}
\vspace{-15mm}
\caption{The Controlled NOT gates of three types ((A), (B), (C) 
from the above) for the three atoms 
in the cavity}
\end{center}
\end{figure}
\vspace{-10mm}

To embed three--qubits in our quantized system we consider 
the simplest one as a wave function 
\begin{equation}
\label{eq:encode three qubits}
\ket{\psi(t)}=
\left(
  \begin{array}{c}
    a_{+++}(t) \\
    a_{++-}(t) \\
    a_{+-+}(t) \\
    a_{+--}(t) \\
    a_{-++}(t) \\
    a_{-+-}(t) \\
    a_{--+}(t) \\
    a_{---}(t) \\
  \end{array}
\right)\otimes \ket{0}.
\end{equation}
like in the case of two--qubits (\ref{eq:encode two qubits}). 

To determine a dynamics that the coefficients 
$a_{+++},a_{++-},\cdots,a_{---}$ satisfy we substitute 
(\ref{eq:encode three qubits}) into the equation 
\begin{eqnarray*}
i\frac{d}{dt}\ket{\psi(t)}
&=&F(t)\ket{\psi(t)}
=\mbox{e}^{itgA_{3}}\tilde{V}(t)\mbox{e}^{-itgA_{3}}\ket{\psi(t)}
    \nonumber \\
&=&
T
\left(
  \begin{array}{ccc}
  \mbox{e}^{itgB_{1/2}} &       &    \\
    & \mbox{e}^{itgB_{1/2}}     &    \\
    &   & \mbox{e}^{itgB_{3/2}}
  \end{array}
\right)
T^{\dagger}\tilde{V}(t)T
\left(
  \begin{array}{ccc}
  \mbox{e}^{-itgB_{1/2}} &       &    \\
    & \mbox{e}^{-itgB_{1/2}}     &    \\
    &   & \mbox{e}^{-itgB_{3/2}}
  \end{array}
\right)
T^{\dagger}\ket{\psi(t)}.
\end{eqnarray*}

Let us rewrite the above equation. If we set 
\begin{equation}
\label{eq:ansatz-modified-three}
\ket{\varphi(t)}\equiv T^{\dagger}\ket{\psi(t)} \Longleftrightarrow 
\left(
  \begin{array}{c}
    \varphi_{0}(t) \\
    \varphi_{1}(t) \\
    \varphi_{2}(t) \\
    \varphi_{3}(t) \\
    \varphi_{4}(t) \\
    \varphi_{5}(t) \\
    \varphi_{6}(t) \\
    \varphi_{7}(t) 
  \end{array}
\right)\otimes \ket{0}
\equiv
\left(
  \begin{array}{c}
    \frac{a_{++-}(t)-a_{+-+}(t)}{\sqrt{2}}           \\
    \frac{a_{-+-}(t)-a_{--+}(t)}{\sqrt{2}}           \\
    \frac{a_{++-}(t)+a_{+-+}(t)-2a_{-++}}{\sqrt{6}}  \\
    \frac{2a_{+--}(t)-a_{-+-}(t)-a_{--+}}{\sqrt{6}}  \\
    a_{+++}                                          \\
    \frac{a_{++-}(t)+a_{+-+}(t)+a_{-++}}{\sqrt{3}}   \\
    \frac{a_{+--}(t)+a_{-+-}(t)+a_{--+}}{\sqrt{3}}   \\
    a_{---}(t)                             
  \end{array}
\right)\otimes \ket{0}
\end{equation}
then we have
\begin{equation}
\label{eq:reduction-equation-three}
i\frac{d}{dt}\ket{\varphi(t)}=
\left(
  \begin{array}{ccc}
  \mbox{e}^{itgB_{1/2}} &       &    \\
    & \mbox{e}^{itgB_{1/2}}     &    \\
    &        & \mbox{e}^{itgB_{3/2}}
  \end{array}
\right)
T^{\dagger}\tilde{V}(t)T
\left(
  \begin{array}{ccc}
  \mbox{e}^{-itgB_{1/2}} &       &    \\
    & \mbox{e}^{-itgB_{1/2}}     &    \\
    &        & \mbox{e}^{-itgB_{3/2}}
  \end{array}
\right)
\ket{\varphi(t)}.
\end{equation}
On the other hand, we have calculated $\mbox{e}^{-itgB_{1/2}}$ in 
(\ref{eq:exponential-1/2}), $\mbox{e}^{-itgB_{3/2}}$ in 
(\ref{eq:exponential-3/2}) and the middle term $T^{\dagger}\tilde{V}(t)T$ in 
(\ref{eq:S-N-V-S-N-three-more}). 

\par \noindent
Therefore the (full) equation is reduced to the equations of 
$\{\varphi_{0},\varphi_{1},\cdots,\varphi_{7}\}$ at the ground state 
$\ket{0}$ : 
\begin{equation}
\label{eq:matrix-form-three}
i\frac{d}{dt}
\left(
  \begin{array}{c}
    \varphi_{0}(t) \\
    \varphi_{1}(t) \\
    \varphi_{2}(t) \\
    \varphi_{3}(t) \\
    \varphi_{4}(t) \\
    \varphi_{5}(t) \\
    \varphi_{6}(t) \\
    \varphi_{7}(t) 
  \end{array}
\right)
=
\left(
  \begin{array}{cccccccc}
    0     & x_{12} &  0     & x_{14} & x_{15} & 0      & x_{17} & 0      \\
   x_{21} &   0    & x_{23} & 0      & 0      & x_{26} & 0      & x_{28} \\
    0     & x_{32} &  0     & x_{34} & x_{35} & 0      & x_{37} & 0      \\
   x_{41} & 0      & x_{43} & 0      & 0      & x_{46} & 0      & x_{48} \\
   x_{51} & 0      & x_{53} & 0      & 0      & x_{56} & 0      & 0      \\
    0     & x_{62} &  0     & x_{64} & x_{65} & 0      & x_{67} & 0      \\
   x_{71} & 0      & x_{73} & 0      & 0      & x_{76} & 0      & x_{78} \\
    0     & x_{82} &  0     & x_{84} & 0      & 0      & x_{87} & 0   
  \end{array}
\right)
\left(
  \begin{array}{c}
    \varphi_{0}(t) \\
    \varphi_{1}(t) \\
    \varphi_{2}(t) \\
    \varphi_{3}(t) \\
    \varphi_{4}(t) \\
    \varphi_{5}(t) \\
    \varphi_{6}(t) \\
    \varphi_{7}(t) 
  \end{array}
\right)
\end{equation}
where 
\begin{eqnarray*}
x_{12}&=&pC(0)C(1), \\
x_{14}&=&\frac{q-r}{\sqrt{3}}C(0)C(1), \\
x_{15}&=&\bar{x}_{51}, \\
x_{17}&=&\frac{q-r}{\sqrt{6}}\left(f_{0}(0)C(1)+3F_{0}(0)S(1)\right), \\
x_{21}&=&\bar{x}_{12}, \\
x_{23}&=&\bar{x}_{32}, \\
x_{26}&=&\bar{x}_{62}, \\
x_{28}&=&\frac{q-r}{\sqrt{2}}f_{-1}(-1)C(0), \\
x_{32}&=&\frac{q-r}{\sqrt{3}}C(0)C(1), \\
x_{34}&=&\frac{-p+2q+2r}{3}C(0)C(1), \\
x_{35}&=&\bar{x}_{53}, \\
x_{37}&=&\frac{2p-q-r}{3\sqrt{2}}\left(f_{0}(0)C(1)+3F_{0}(0)S(1)\right), \\
x_{41}&=&\bar{x}_{14}, \\
x_{43}&=&\bar{x}_{34}, \\
x_{46}&=&\bar{x}_{64}, \\
x_{48}&=&\frac{2p-q-r}{\sqrt{6}}f_{-1}(-1)C(0), \\
x_{51}&=&\frac{-q+r}{\sqrt{2}}\left(C(1)f_{2}(2)+S(1)F_{1}(2)\right), \\
x_{53}&=&\frac{-2p+q+r}{\sqrt{6}}\left(C(1)f_{2}(2)+S(1)F_{1}(2)\right), \\
x_{56}&=&\frac{p+q+r}{\sqrt{3}}\left(f_{1}(1)f_{2}(2)+4H_{1}(1)F_{1}(2)+
24h_{1}(1)h_{1}(2)\right), \\
x_{62}&=&\frac{-q+r}{\sqrt{2}}C(0)f_{1}(1), \\
x_{64}&=&\frac{-2p+q+r}{3\sqrt{2}}C(0)f_{1}(1), \\
x_{65}&=&\bar{x}_{56}, \\
x_{67}&=&\frac{2(p+q+r)}{3}\left(f_{0}(0)f_{1}(1)+3F_{0}(0)H_{1}(1)
\right), \\
x_{71}&=&\bar{x}_{17}, \\
x_{73}&=&\bar{x}_{37}, \\
x_{76}&=&\bar{x}_{67}, \\
x_{78}&=&\frac{p+q+r}{\sqrt{3}}f_{-1}(-1)f_{0}(0), \\
x_{82}&=&\bar{x}_{28}, \\
x_{84}&=&\bar{x}_{48}, \\
x_{87}&=&\bar{x}_{78}
\end{eqnarray*}
and $C(0)$, $C(1)$, $S(1)$, $f_{-1}(-1)$, $f_{0}(0)$, $f_{1}(1)$, $f_{2}(2)$, 
$F_{0}(0)$, $F_{1}(2)$, $h_{1}(1)$, $h_{1}(2)$, $H_{1}(1)$ are respectively 
given as 
\begin{eqnarray*}
p(t)&=&h_{1}\mbox{e}^{i\{(\Omega_{1}+\omega)t+\phi_{1}\}},\quad 
q(t)=h_{2}\mbox{e}^{i\{(\Omega_{2}+\omega)t+\phi_{2}\}},\quad
r(t)=h_{3}\mbox{e}^{i\{(\Omega_{3}+\omega)t+\phi_{3}\}}, \\
C(0)&=&1, \\
C(1)&=&\mbox{cos}(tg), \\
S(1)&=&\mbox{sin}(tg), \\
f_{-1}(-1)&=&1, \\
f_{0}(0)&=&\mbox{cos}(tg\sqrt{3}), \\
f_{1}(1)&=&\frac{2\mbox{cos}(tg\sqrt{10})+3}{5}, \\
f_{2}(2)&=&\frac{(-7+\sqrt{73})\mbox{cos}(tg\sqrt{10+\sqrt{73}})+
(7+\sqrt{73})\mbox{cos}(tg\sqrt{10-\sqrt{73}})}{2\sqrt{73}}, \\
F_{0}(0)&=&\frac{\mbox{sin}(tg\sqrt{3})}{\sqrt{3}}, \\
F_{1}(2)&=&\frac{1}{2\sqrt{73}}
\left\{\frac{1+\sqrt{73}}{\sqrt{10+\sqrt{73}}}
\mbox{sin}(tg\sqrt{10+\sqrt{73}})-\frac{1-\sqrt{73}}{\sqrt{10-\sqrt{73}}}
\mbox{sin}(tg\sqrt{10-\sqrt{73}})\right\}, \\
f_{3}(1)&=&\frac{\mbox{cos}(tg\sqrt{10})-1}{10}, \\
f_{3}(2)&=&\frac{\mbox{cos}(tg\sqrt{10+\sqrt{73}})-
\mbox{cos}(tg\sqrt{10-\sqrt{73}})}{2\sqrt{73}}, \\
H_{1}(1)&=&\frac{\mbox{sin}(tg\sqrt{10})}{\sqrt{10}}.
\end{eqnarray*}
These equations are complicated enough. 

Next let us derive the three controlled NOT gates (A), (B), (C) in Figure 2 
from the matrix equation above. For that we use a resonance condition and 
rotating wave approximation associated to it.

\par \vspace{5mm} \noindent
{\bf Derivation of (A)}

We focus on the components $x_{56}$ and $x_{65}$ = $\bar{x}_{56}$ 
in the matrix. $f_{1}(1)f_{2}(2)$ and $h_{1}(1)h_{1}(2)$ 
contain the term $\mbox{e}^{-itg\sqrt{10+\sqrt{73}}}$ coming from the Euler 
formula, and therefore $x_{56}$ contain the oscillating term \footnote
{We use $p(t)$ in the first atom, however it is of course possible to use 
$q(t)$ (the second atom) or $r(t)$ (the third atom)}
\[
\mbox{e}^{i\{(\Omega_{1}+\omega)t+\phi_{1}\}}
\mbox{e}^{-itg\sqrt{10+\sqrt{73}}}
=\mbox{e}^{i\{(\Omega_{1}+\omega-g\sqrt{10+\sqrt{73}})t+\phi_{1}\}}.
\]
We note that this term is not contained in other components in the matrix. 
Here we set the resonance condition 
\begin{equation}
\label{eq:resonance-(A)}
\Omega_{1}+\omega-g\sqrt{10+\sqrt{73}}=0
\end{equation}
and apply the rotating wave approximation associated to this. 
Then the above complicated matrix equation becomes a very simple one 

\begin{equation}
\label{eq:reduced-matrix-form-(A)}
i\frac{d}{dt}
\left(
  \begin{array}{c}
    \varphi_{0}(t) \\
    \varphi_{1}(t) \\
    \varphi_{2}(t) \\
    \varphi_{3}(t) \\
    \varphi_{4}(t) \\
    \varphi_{5}(t) \\
    \varphi_{6}(t) \\
    \varphi_{7}(t) 
  \end{array}
\right)
=
\frac{\sqrt{3}(-11+\sqrt{73})h_{1}}{20\sqrt{73}}
\left(
  \begin{array}{cccccccc}
    0 & 0 & 0 & 0 & 0 & 0 & 0 & 0                     \\
    0 & 0 & 0 & 0 & 0 & 0 & 0 & 0                     \\
    0 & 0 & 0 & 0 & 0 & 0 & 0 & 0                     \\
    0 & 0 & 0 & 0 & 0 & 0 & 0 & 0                     \\
    0 & 0 & 0 & 0 & 0 & \mbox{e}^{i\phi_{1}} & 0 & 0  \\
    0 & 0 & 0 & 0 & \mbox{e}^{-i\phi_{1}} & 0 & 0 & 0 \\
    0 & 0 & 0 & 0 & 0 & 0 & 0 & 0                     \\
    0 & 0 & 0 & 0 & 0 & 0 & 0 & 0 
  \end{array}
\right)
\left(
  \begin{array}{c}
    \varphi_{0}(t) \\
    \varphi_{1}(t) \\
    \varphi_{2}(t) \\
    \varphi_{3}(t) \\
    \varphi_{4}(t) \\
    \varphi_{5}(t) \\
    \varphi_{6}(t) \\
    \varphi_{7}(t) 
  \end{array}
\right).
\end{equation}

The solution is easily obtained to be
\begin{eqnarray}
\label{eq:solution-(A)}
\left(
  \begin{array}{c}
    \varphi_{0}(t) \\
    \varphi_{1}(t) \\
    \varphi_{2}(t) \\
    \varphi_{3}(t) \\
    \varphi_{4}(t) \\
    \varphi_{5}(t) \\
    \varphi_{6}(t) \\
    \varphi_{7}(t) 
  \end{array}
\right)
&=&
\mbox{exp}
\left\{it
\frac{\sqrt{3}(11-\sqrt{73})h_{1}}{20\sqrt{73}}
\left(
  \begin{array}{cccccccc}
    0 & 0 & 0 & 0 & 0 & 0 & 0 & 0                     \\
    0 & 0 & 0 & 0 & 0 & 0 & 0 & 0                     \\
    0 & 0 & 0 & 0 & 0 & 0 & 0 & 0                     \\
    0 & 0 & 0 & 0 & 0 & 0 & 0 & 0                     \\
    0 & 0 & 0 & 0 & 0 & \mbox{e}^{i\phi_{1}} & 0 & 0  \\
    0 & 0 & 0 & 0 & \mbox{e}^{-i\phi_{1}} & 0 & 0 & 0 \\
    0 & 0 & 0 & 0 & 0 & 0 & 0 & 0                     \\
    0 & 0 & 0 & 0 & 0 & 0 & 0 & 0 
  \end{array}
\right)
\right\}
\left(
  \begin{array}{c}
    \varphi_{0}(0) \\
    \varphi_{1}(0) \\
    \varphi_{2}(0) \\
    \varphi_{3}(0) \\
    \varphi_{4}(0) \\
    \varphi_{5}(0) \\
    \varphi_{6}(0) \\
    \varphi_{7}(0) 
  \end{array}
\right)                \nonumber \\
&=&
\left(
  \begin{array}{cccccccc}
    1 & 0 & 0 & 0 & 0 & 0 & 0 & 0                               \\
    0 & 1 & 0 & 0 & 0 & 0 & 0 & 0                               \\
    0 & 0 & 1 & 0 & 0 & 0 & 0 & 0                               \\
    0 & 0 & 0 & 1 & 0 & 0 & 0 & 0                               \\
    0 & 0 & 0 & 0 & \mbox{cos}(\alpha t) & 
    i\mbox{e}^{i\phi_{1}}\mbox{sin}(\alpha t) & 0 & 0           \\
    0 & 0 & 0 & 0 & i\mbox{e}^{-i\phi_{1}}\mbox{sin}(\alpha t)
    & \mbox{cos}(\alpha t) & 0 & 0                              \\
    0 & 0 & 0 & 0 & 0 & 0 & 1 & 0                               \\
    0 & 0 & 0 & 0 & 0 & 0 & 0 & 1 
  \end{array}
\right)
\left(
  \begin{array}{c}
    \varphi_{0}(0) \\
    \varphi_{1}(0) \\
    \varphi_{2}(0) \\
    \varphi_{3}(0) \\
    \varphi_{4}(0) \\
    \varphi_{5}(0) \\
    \varphi_{6}(0) \\
    \varphi_{7}(0) 
  \end{array}
\right)                   \nonumber \\
&\equiv&
U_{A}(t)
\left(
  \begin{array}{c}
    \varphi_{0}(0) \\
    \varphi_{1}(0) \\
    \varphi_{2}(0) \\
    \varphi_{3}(0) \\
    \varphi_{4}(0) \\
    \varphi_{5}(0) \\
    \varphi_{6}(0) \\
    \varphi_{7}(0) 
  \end{array}
\right)
\end{eqnarray}
where we have set $\alpha=\frac{\sqrt{3}(11-\sqrt{73})h_{1}}{20\sqrt{73}}$.

Here if we choose $t_{A}$ as $\mbox{cos}(\alpha t_{A})=-1$, 
then
\[
U_{A}(t_{A})
=
\left(
  \begin{array}{cccccccc}
    1 & 0 & 0 & 0 & 0 & 0 & 0 & 0    \\
    0 & 1 & 0 & 0 & 0 & 0 & 0 & 0    \\
    0 & 0 & 1 & 0 & 0 & 0 & 0 & 0    \\
    0 & 0 & 0 & 1 & 0 & 0 & 0 & 0    \\
    0 & 0 & 0 & 0 &-1 & 0 & 0 & 0    \\
    0 & 0 & 0 & 0 & 0 &-1 & 0 & 0    \\
    0 & 0 & 0 & 0 & 0 & 0 & 1 & 0    \\
    0 & 0 & 0 & 0 & 0 & 0 & 0 & 1 
  \end{array}
\right)
=
\left(
  \begin{array}{cccc}
     1 &     &     &     \\
       &  1  &     &     \\
       &     & -1  &     \\
       &     &     &  1
  \end{array}
\right)
\otimes
\left(
  \begin{array}{cc}
     1 &     \\
       &  1  \\
  \end{array}
\right).
\]
By multiplying ${\bf 1}_{2}\otimes \sigma_{1}\otimes {\bf 1}_{2}$ from both 
sides we have 
\begin{equation}
\label{eq:special-A}
\tilde{U}_{A}(t_{A})\equiv
({\bf 1}_{2}\otimes \sigma_{1}\otimes {\bf 1}_{2})
U_{A}(t_{A})
({\bf 1}_{2}\otimes \sigma_{1}\otimes {\bf 1}_{2}) \\
=
\left(
  \begin{array}{cccc}
     1 &     &     &     \\
       &  1  &     &     \\
       &     &  1  &     \\
       &     &     &  -1
  \end{array}
\right)
\otimes
\left(
  \begin{array}{cc}
     1 &     \\
       &  1  \\
  \end{array}
\right).
\end{equation}

Moreover, by multiplying ${\bf 1}_{2}\otimes W\otimes {\bf 1}_{2}$ from both 
sides we finally obtain the case (A) in Figure 2 
\begin{equation}
({\bf 1}_{2}\otimes W\otimes {\bf 1}_{2})
\tilde{U}_{A}(t_{A})
({\bf 1}_{2}\otimes W\otimes {\bf 1}_{2})
=
\left(
  \begin{array}{cccc}
     1 &     &     &     \\
       &  1  &     &     \\
       &     &  0  &  1  \\
       &     &  1  &  0
  \end{array}
\right)
\otimes
\left(
  \begin{array}{cc}
     1 &     \\
       &  1  \\
  \end{array}
\right)
=C_{NOT}\otimes {\bf 1}_{2}.
\end{equation}

\par \vspace{5mm} \noindent
{\bf Derivation of (B)}

We focus on the components $x_{37}$ and $x_{73}$ = $\bar{x}_{37}$ 
in the matrix. $f_{0}(0)C(1)$ and $F_{0}(0)S(1)$ 
contain the term $\mbox{e}^{-itg(1+\sqrt{3})}$ coming from the Euler 
formula, and therefore $x_{37}$ contain the oscillating term 
\[
\mbox{e}^{i\{(\Omega_{1}+\omega)t+\phi_{1}\}}
\mbox{e}^{-itg(1+\sqrt{3})}
=\mbox{e}^{i\{\left(\Omega_{1}+\omega-g(1+\sqrt{3})\right)t+\phi_{1}\}}.
\]
This term is not contained in other components in the matrix. 
Here we set the resonance condition 
\begin{equation}
\label{eq:resonance-(B)}
\Omega_{1}+\omega-g(1+\sqrt{3})=0
\end{equation}
and apply the rotating wave approximation associated to this. 
Then the above complicated matrix equation becomes a very simple one 

\begin{equation}
\label{eq:reduced-matrix-form-(B)}
i\frac{d}{dt}
\left(
  \begin{array}{c}
    \varphi_{0}(t) \\
    \varphi_{1}(t) \\
    \varphi_{2}(t) \\
    \varphi_{3}(t) \\
    \varphi_{4}(t) \\
    \varphi_{5}(t) \\
    \varphi_{6}(t) \\
    \varphi_{7}(t) 
  \end{array}
\right)
=
\frac{\sqrt{2}(1-\sqrt{3})h_{1}}{12}
\left(
  \begin{array}{cccccccc}
    0 & 0 & 0 & 0 & 0 & 0 & 0 & 0                     \\
    0 & 0 & 0 & 0 & 0 & 0 & 0 & 0                     \\
    0 & 0 & 0 & 0 & 0 & 0 & \mbox{e}^{i\phi_{1}} & 0  \\
    0 & 0 & 0 & 0 & 0 & 0 & 0 & 0                     \\
    0 & 0 & 0 & 0 & 0 & 0 & 0 & 0                     \\
    0 & 0 & 0 & 0 & 0 & 0 & 0 & 0                     \\
    0 & 0 & \mbox{e}^{-i\phi_{1}} & 0 & 0 & 0 & 0 & 0 \\
    0 & 0 & 0 & 0 & 0 & 0 & 0 & 0 
  \end{array}
\right)
\left(
  \begin{array}{c}
    \varphi_{0}(t) \\
    \varphi_{1}(t) \\
    \varphi_{2}(t) \\
    \varphi_{3}(t) \\
    \varphi_{4}(t) \\
    \varphi_{5}(t) \\
    \varphi_{6}(t) \\
    \varphi_{7}(t) 
  \end{array}
\right).
\end{equation}

The solution is easily obtained to be
\begin{eqnarray}
\label{eq:solution-(B)}
\left(
  \begin{array}{c}
    \varphi_{0}(t) \\
    \varphi_{1}(t) \\
    \varphi_{2}(t) \\
    \varphi_{3}(t) \\
    \varphi_{4}(t) \\
    \varphi_{5}(t) \\
    \varphi_{6}(t) \\
    \varphi_{7}(t) 
  \end{array}
\right)
&=&
\mbox{exp}
\left\{it
\frac{\sqrt{2}(-1+\sqrt{3})h_{1}}{12}
\left(
  \begin{array}{cccccccc}
    0 & 0 & 0 & 0 & 0 & 0 & 0 & 0                     \\
    0 & 0 & 0 & 0 & 0 & 0 & 0 & 0                     \\
    0 & 0 & 0 & 0 & 0 & 0 & \mbox{e}^{i\phi_{1}} & 0  \\
    0 & 0 & 0 & 0 & 0 & 0 & 0 & 0                     \\
    0 & 0 & 0 & 0 & 0 & 0 & 0 & 0                     \\
    0 & 0 & 0 & 0 & 0 & 0 & 0 & 0                     \\
    0 & 0 & \mbox{e}^{-i\phi_{1}} & 0 & 0 & 0 & 0 & 0 \\
    0 & 0 & 0 & 0 & 0 & 0 & 0 & 0 
  \end{array}
\right)
\right\}
\left(
  \begin{array}{c}
    \varphi_{0}(0) \\
    \varphi_{1}(0) \\
    \varphi_{2}(0) \\
    \varphi_{3}(0) \\
    \varphi_{4}(0) \\
    \varphi_{5}(0) \\
    \varphi_{6}(0) \\
    \varphi_{7}(0) 
  \end{array}
\right)                \nonumber \\
&=&
\left(
  \begin{array}{cccccccc}
    1 & 0 & 0 & 0 & 0 & 0 & 0 & 0                               \\
    0 & 1 & 0 & 0 & 0 & 0 & 0 & 0                               \\
    0 & 0 & \mbox{cos}(\beta t) & 0 & 0 & 0 & 
    i\mbox{e}^{i\phi_{1}}\mbox{sin}(\beta t) & 0                \\
    0 & 0 & 0 & 1 & 0 & 0 & 0 & 0                               \\
    0 & 0 & 0 & 0 & 1 & 0 & 0 & 0                               \\
    0 & 0 & 0 & 0 & 0 & 1 & 0 & 0                               \\
    0 & 0 & i\mbox{e}^{-i\phi_{1}}\mbox{sin}(\beta t) & 0 & 0 & 
    0 & \mbox{cos}(\beta t) & 0                                 \\
    0 & 0 & 0 & 0 & 0 & 0 & 0 & 1 
  \end{array}
\right)
\left(
  \begin{array}{c}
    \varphi_{0}(0) \\
    \varphi_{1}(0) \\
    \varphi_{2}(0) \\
    \varphi_{3}(0) \\
    \varphi_{4}(0) \\
    \varphi_{5}(0) \\
    \varphi_{6}(0) \\
    \varphi_{7}(0) 
  \end{array}
\right)                   \nonumber \\
&\equiv&
U_{B}(t)
\left(
  \begin{array}{c}
    \varphi_{0}(0) \\
    \varphi_{1}(0) \\
    \varphi_{2}(0) \\
    \varphi_{3}(0) \\
    \varphi_{4}(0) \\
    \varphi_{5}(0) \\
    \varphi_{6}(0) \\
    \varphi_{7}(0) 
  \end{array}
\right)
\end{eqnarray}
where we have set $\beta=\frac{\sqrt{2}(-1+\sqrt{3})h_{1}}{12}$.

Here if we choose $t_{B}$ as $\mbox{cos}(\beta t_{B})=-1$, 
then
\[
U_{B}(t_{B})
=
\left(
  \begin{array}{cccccccc}
    1 & 0 & 0 & 0 & 0 & 0 & 0 & 0    \\
    0 & 1 & 0 & 0 & 0 & 0 & 0 & 0    \\
    0 & 0 &-1 & 0 & 0 & 0 & 0 & 0    \\
    0 & 0 & 0 & 1 & 0 & 0 & 0 & 0    \\
    0 & 0 & 0 & 0 & 1 & 0 & 0 & 0    \\
    0 & 0 & 0 & 0 & 0 & 1 & 0 & 0    \\
    0 & 0 & 0 & 0 & 0 & 0 &-1 & 0    \\
    0 & 0 & 0 & 0 & 0 & 0 & 0 & 1 
  \end{array}
\right)
=
\left(
  \begin{array}{cc}
     1 &     \\
       &  1  \\
  \end{array}
\right)
\otimes
\left(
  \begin{array}{cccc}
     1 &     &     &     \\
       &  1  &     &     \\
       &     & -1  &     \\
       &     &     &  1
  \end{array}
\right).
\]

By multiplying ${\bf 1}_{2}\otimes {\bf 1}_{2}\otimes \sigma_{1}$ from both 
sides we have 
\[
\tilde{U}_{B}(t_{B})\equiv
({\bf 1}_{2}\otimes {\bf 1}_{2}\otimes \sigma_{1})
U_{B}(t_{B})
({\bf 1}_{2}\otimes {\bf 1}_{2}\otimes \sigma_{1})
 \\
=
{\bf 1}_{2}\otimes
\left(
  \begin{array}{cccc}
     1 &     &     &     \\
       &  1  &     &     \\
       &     &  1  &     \\
       &     &     &  -1
  \end{array}
\right).
\]

Moreover, by multiplying ${\bf 1}_{2}\otimes {\bf 1}_{2}\otimes W$ from both 
sides we finally obtain the case (B) in Figure 2 
\begin{equation}
({\bf 1}_{2}\otimes {\bf 1}_{2}\otimes W)
\tilde{U}_{B}(t_{B})
({\bf 1}_{2}\otimes {\bf 1}_{2}\otimes W)
=
{\bf 1}_{2}\otimes
\left(
  \begin{array}{cccc}
     1 &     &     &     \\
       &  1  &     &     \\
       &     &  0  &  1  \\
       &     &  1  &  0
  \end{array}
\right)
={\bf 1}_{2}\otimes C_{NOT}.
\end{equation}

\par \vspace{5mm} \noindent
{\bf Derivation of (C)}

It is well--known that the case (C) is obtained as 

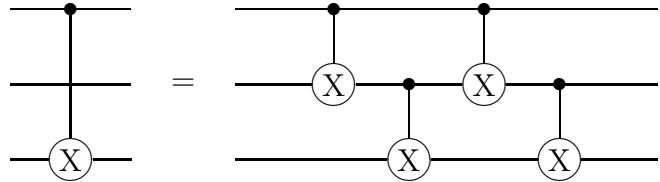
\begin{center}
\setlength{\unitlength}{1mm} 
\begin{picture}(110,40)(0,-10)
\put(10,20){\line(1,0){16}}
\put(10,10){\line(1,0){16}}
\put(10, 0){\line(1,0){5}}
\put(21, 0){\line(1,0){5}}
\put(15,17){\makebox(6,6){$\bullet$}}
\put(18, 0){\circle{6}}  
\put(15,-3){\makebox(6,6){X}}
\put(18,20){\line(0,-1){17}} 
\put(30,7){\makebox(6,6){$=$}} 
\put(40,20){\line(1,0){56}}
\put(40, 0){\line(1,0){20}}
\put(66, 0){\line(1,0){14}}
\put(86, 0){\line(1,0){10}}
\put(40,10){\line(1,0){10}}
\put(56,10){\line(1,0){14}}
\put(76,10){\line(1,0){20}}
\put(50,17){\makebox(6,6){$\bullet$}}
\put(53,10){\circle{6}}  
\put(50, 7){\makebox(6,6){X}}
\put(53,20){\line(0,-1){7}}
\put(70,17){\makebox(6,6){$\bullet$}}
\put(73,10){\circle{6}}  
\put(70, 7){\makebox(6,6){X}}
\put(73,20){\line(0,-1){7}}
\put(60, 7){\makebox(6,6){$\bullet$}}
\put(63, 0){\circle{6}}  
\put(60,-3){\makebox(6,6){X}}
\put(63,10){\line(0,-1){7}}
\put(80, 7){\makebox(6,6){$\bullet$}}
\put(83, 0){\circle{6}}  
\put(80,-3){\makebox(6,6){X}}
\put(83,10){\line(0,-1){7}}
\end{picture}
\end{center}
%
by making use of (A) and (B). However, 4 steps are needed in construction. 
We give a direct construction like (A) or (B) which is favorable from the 
point of view of quick construction.

We focus on the components $x_{67}$ and $x_{76}$ = $\bar{x}_{67}$ 
in the matrix. $f_{0}(0)f_{1}(1)$ and $F_{0}(0)H_{1}(1)$ 
contain the term $\mbox{e}^{-itg(\sqrt{3}+\sqrt{10})}$ coming from the Euler 
formula, and therefore $x_{67}$ contain the oscillating term 
\[
\mbox{e}^{i\{(\Omega_{1}+\omega)t+\phi_{1}\}}
\mbox{e}^{-itg(\sqrt{3}+\sqrt{10})}
=\mbox{e}^{i\{\left(\Omega_{1}+\omega-g(\sqrt{3}+\sqrt{10})\right)t+
\phi_{1}\}}.
\]
This term is not contained in other components in the matrix. 
Here we set the resonance condition 
\begin{equation}
\label{eq:resonance-(C)}
\Omega_{1}+\omega-g(\sqrt{3}+\sqrt{10})=0
\end{equation}
and apply the rotating wave approximation associated to this. 
Then the above complicated matrix equation becomes a very simple one 

\begin{equation}
\label{eq:reduced-matrix-form-(C)}
i\frac{d}{dt}
\left(
  \begin{array}{c}
    \varphi_{0}(t) \\
    \varphi_{1}(t) \\
    \varphi_{2}(t) \\
    \varphi_{3}(t) \\
    \varphi_{4}(t) \\
    \varphi_{5}(t) \\
    \varphi_{6}(t) \\
    \varphi_{7}(t) 
  \end{array}
\right)
=
\frac{(4-\sqrt{30})h_{1}}{60}
\left(
  \begin{array}{cccccccc}
    0 & 0 & 0 & 0 & 0 & 0 & 0 & 0                     \\
    0 & 0 & 0 & 0 & 0 & 0 & 0 & 0                     \\
    0 & 0 & 0 & 0 & 0 & 0 & 0 & 0                     \\
    0 & 0 & 0 & 0 & 0 & 0 & 0 & 0                     \\
    0 & 0 & 0 & 0 & 0 & 0 & 0 & 0                     \\
    0 & 0 & 0 & 0 & 0 & 0 & \mbox{e}^{i\phi_{1}} & 0  \\
    0 & 0 & 0 & 0 & 0 & \mbox{e}^{-i\phi_{1}} & 0 & 0 \\
    0 & 0 & 0 & 0 & 0 & 0 & 0 & 0 
  \end{array}
\right)
\left(
  \begin{array}{c}
    \varphi_{0}(t) \\
    \varphi_{1}(t) \\
    \varphi_{2}(t) \\
    \varphi_{3}(t) \\
    \varphi_{4}(t) \\
    \varphi_{5}(t) \\
    \varphi_{6}(t) \\
    \varphi_{7}(t) 
  \end{array}
\right).
\end{equation}

The solution is easily obtained to be
\begin{eqnarray}
\label{eq:solution-(C)}
\left(
  \begin{array}{c}
    \varphi_{0}(t) \\
    \varphi_{1}(t) \\
    \varphi_{2}(t) \\
    \varphi_{3}(t) \\
    \varphi_{4}(t) \\
    \varphi_{5}(t) \\
    \varphi_{6}(t) \\
    \varphi_{7}(t) 
  \end{array}
\right)
&=&
\mbox{exp}
\left\{it
\frac{(-4+\sqrt{30})h_{1}}{60}
\left(
  \begin{array}{cccccccc}
    0 & 0 & 0 & 0 & 0 & 0 & 0 & 0                     \\
    0 & 0 & 0 & 0 & 0 & 0 & 0 & 0                     \\
    0 & 0 & 0 & 0 & 0 & 0 & 0 & 0                     \\
    0 & 0 & 0 & 0 & 0 & 0 & 0 & 0                     \\
    0 & 0 & 0 & 0 & 0 & 0 & 0 & 0                     \\
    0 & 0 & 0 & 0 & 0 & 0 & \mbox{e}^{i\phi_{1}} & 0  \\
    0 & 0 & 0 & 0 & 0 & \mbox{e}^{-i\phi_{1}} & 0 & 0 \\
    0 & 0 & 0 & 0 & 0 & 0 & 0 & 0 
  \end{array}
\right)
\right\}
\left(
  \begin{array}{c}
    \varphi_{0}(0) \\
    \varphi_{1}(0) \\
    \varphi_{2}(0) \\
    \varphi_{3}(0) \\
    \varphi_{4}(0) \\
    \varphi_{5}(0) \\
    \varphi_{6}(0) \\
    \varphi_{7}(0) 
  \end{array}
\right)                \nonumber \\
&=&
\left(
  \begin{array}{cccccccc}
    1 & 0 & 0 & 0 & 0 & 0 & 0 & 0                                  \\
    0 & 1 & 0 & 0 & 0 & 0 & 0 & 0                                  \\
    0 & 0 & 1 & 0 & 0 & 0 & 0 & 0                                  \\
    0 & 0 & 0 & 1 & 0 & 0 & 0 & 0                                  \\
    0 & 0 & 0 & 0 & 1 & 0 & 0 & 0                                  \\
    0 & 0 & 0 & 0 & 0 & \mbox{cos}(\gamma t) & 
    i\mbox{e}^{i\phi_{1}}\mbox{sin}(\gamma t) & 0                  \\
    0 & 0 & 0 & 0 & 0 & i\mbox{e}^{-i\phi_{1}}\mbox{sin}(\gamma t) 
    & \mbox{cos}(\gamma t) & 0                                     \\
    0 & 0 & 0 & 0 & 0 & 0 & 0 & 1 
  \end{array}
\right)
\left(
  \begin{array}{c}
    \varphi_{0}(0) \\
    \varphi_{1}(0) \\
    \varphi_{2}(0) \\
    \varphi_{3}(0) \\
    \varphi_{4}(0) \\
    \varphi_{5}(0) \\
    \varphi_{6}(0) \\
    \varphi_{7}(0) 
  \end{array}
\right)                   \nonumber \\
&\equiv&
U_{C}(t)
\left(
  \begin{array}{c}
    \varphi_{0}(0) \\
    \varphi_{1}(0) \\
    \varphi_{2}(0) \\
    \varphi_{3}(0) \\
    \varphi_{4}(0) \\
    \varphi_{5}(0) \\
    \varphi_{6}(0) \\
    \varphi_{7}(0) 
  \end{array}
\right)
\end{eqnarray}
where we have set $\gamma=\frac{(-4+\sqrt{30})h_{1}}{60}$.

Here if we choose $t_{C}$ as $\mbox{cos}(\gamma t_{C})=-1$, 
then
\[
U_{C}(t_{C})
=
\left(
  \begin{array}{cccccccc}
    1 & 0 & 0 & 0 & 0 & 0 & 0 & 0    \\
    0 & 1 & 0 & 0 & 0 & 0 & 0 & 0    \\
    0 & 0 & 1 & 0 & 0 & 0 & 0 & 0    \\
    0 & 0 & 0 & 1 & 0 & 0 & 0 & 0    \\
    0 & 0 & 0 & 0 & 1 & 0 & 0 & 0    \\
    0 & 0 & 0 & 0 & 0 &-1 & 0 & 0    \\
    0 & 0 & 0 & 0 & 0 & 0 &-1 & 0    \\
    0 & 0 & 0 & 0 & 0 & 0 & 0 & 1 
  \end{array}
\right).
\]

By multiplying $\tilde{U}_{A}(t_{A})$ in (\ref{eq:special-A}) 
from the left hand side we have 
\[
\tilde{U}_{C}(t_{C})\equiv \tilde{U}_{A}(t_{A})U_{C}(t_{C})
=
\left(
  \begin{array}{cccccccc}
    1 & 0 & 0 & 0 & 0 & 0 & 0 & 0    \\
    0 & 1 & 0 & 0 & 0 & 0 & 0 & 0    \\
    0 & 0 & 1 & 0 & 0 & 0 & 0 & 0    \\
    0 & 0 & 0 & 1 & 0 & 0 & 0 & 0    \\
    0 & 0 & 0 & 0 & 1 & 0 & 0 & 0    \\
    0 & 0 & 0 & 0 & 0 &-1 & 0 & 0    \\
    0 & 0 & 0 & 0 & 0 & 0 & 1 & 0    \\
    0 & 0 & 0 & 0 & 0 & 0 & 0 &-1 
  \end{array}
\right).
\]

Moreover, by multiplying ${\bf 1}_{2}\otimes {\bf 1}_{2}\otimes W$ from both 
sides we finally obtain the case (C) in Figure 2 
\begin{equation}
({\bf 1}_{2}\otimes {\bf 1}_{2}\otimes W)
\tilde{U}_{C}(t_{C})
({\bf 1}_{2}\otimes {\bf 1}_{2}\otimes W)
=
\left(
  \begin{array}{cccccccc}
    1 & 0 & 0 & 0 & 0 & 0 & 0 & 0    \\
    0 & 1 & 0 & 0 & 0 & 0 & 0 & 0    \\
    0 & 0 & 1 & 0 & 0 & 0 & 0 & 0    \\
    0 & 0 & 0 & 1 & 0 & 0 & 0 & 0    \\
    0 & 0 & 0 & 0 & 0 & 1 & 0 & 0    \\
    0 & 0 & 0 & 0 & 1 & 0 & 0 & 0    \\
    0 & 0 & 0 & 0 & 0 & 0 & 0 & 1    \\
    0 & 0 & 0 & 0 & 0 & 0 & 1 & 0 
  \end{array}
\right)
=\tilde{C}_{NOT}.
\end{equation}

\section{Further Problem}

In this section we consider the case of four atoms (the system of four 
qubits) and present a problem on constructing the 
controlled--controlled--controlled NOT gate defined by
%
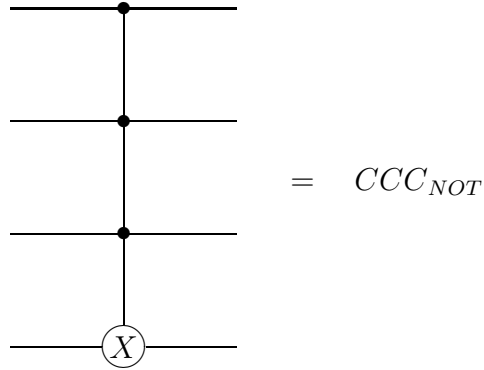
\begin{center}
\setlength{\unitlength}{1mm}  
\begin{picture}(70,60)
\put(10,55){\line(1,0){30}}  
\put(10,40){\line(1,0){30}}   
\put(10,25){\line(1,0){30}} 
\put(10,10){\line(1,0){12}} 
\put(28,10){\line(1,0){12}} 
\put(22,50){\makebox(6,10){$\bullet$}}
\put(22,35){\makebox(6,10){$\bullet$}}
\put(22,20){\makebox(6,10){$\bullet$}}
\put(25,10){\circle{6}}  
\put(22,5){\makebox(6,10){$X$}} 
\put(25,55){\line(0,-1){15}}  
\put(25,40){\line(0,-1){15}}  
\put(25,25){\line(0,-1){12}}  
\put(50,27){\makebox(20,10){$=\quad {CCC}_{NOT}$}}
\end{picture}
\end{center}
\vspace{-5mm}
as a picture. This gate is usually constructed as 

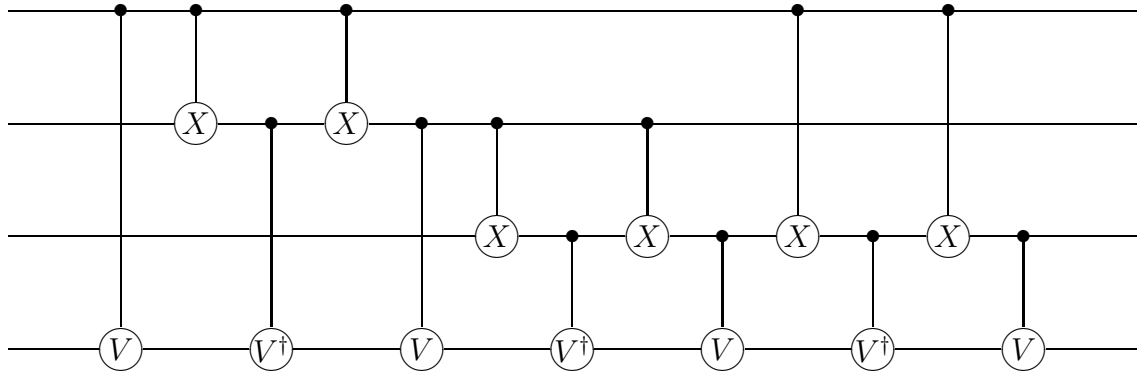
\begin{center}
\setlength{\unitlength}{1mm}  
\begin{picture}(180,60)(10,0)
\put(10,55){\line(1,0){150}}
\put(10,40){\line(1,0){22}}
\put(38,40){\line(1,0){14}}
\put(58,40){\line(1,0){102}}
\put(10,25){\line(1,0){62}} 
\put(78,25){\line(1,0){14}}
\put(98,25){\line(1,0){14}}
\put(118,25){\line(1,0){14}}
\put(138,25){\line(1,0){22}}
\put(10,10){\line(1,0){12}} 
\put(28,10){\line(1,0){14}}
\put(48,10){\line(1,0){14}}
\put(68,10){\line(1,0){14}}
\put(88,10){\line(1,0){14}}
\put(108,10){\line(1,0){14}}
\put(128,10){\line(1,0){14}}
\put(148,10){\line(1,0){12}}
\put(22,50){\makebox(6,10){$\bullet$}}
\put(22, 5){\makebox(6,10){$V$}} 
\put(25,10){\circle{6}} 
\put(25,55){\line(0,-1){42}}  
\put(32,50){\makebox(6,10){$\bullet$}}
\put(32,35){\makebox(6,10){$X$}} 
\put(35,40){\circle{6}} 
\put(35,55){\line(0,-1){12}}  
\put(42,35){\makebox(6,10){$\bullet$}}
\put(42, 5){\makebox(6,10){$V^{\dagger}$}} 
\put(45,10){\circle{6}} 
\put(45,40){\line(0,-1){27}}  
\put(52,50){\makebox(6,10){$\bullet$}}
\put(52,35){\makebox(6,10){$X$}} 
\put(55,40){\circle{6}} 
\put(55,55){\line(0,-1){12}} 
\put(62,35){\makebox(6,10){$\bullet$}}
\put(62, 5){\makebox(6,10){$V$}} 
\put(65,10){\circle{6}} 
\put(65,40){\line(0,-1){27}}  
\put(72,35){\makebox(6,10){$\bullet$}}
\put(72,20){\makebox(6,10){$X$}} 
\put(75,25){\circle{6}} 
\put(75,40){\line(0,-1){12}}  
\put(82,20){\makebox(6,10){$\bullet$}}
\put(82, 5){\makebox(6,10){$V^{\dagger}$}} 
\put(85,10){\circle{6}} 
\put(85,25){\line(0,-1){12}}
\put(92,35){\makebox(6,10){$\bullet$}}
\put(92,20){\makebox(6,10){$X$}} 
\put(95,25){\circle{6}} 
\put(95,40){\line(0,-1){12}}  
\put(102,20){\makebox(6,10){$\bullet$}}
\put(102, 5){\makebox(6,10){$V$}} 
\put(105,10){\circle{6}} 
\put(105,25){\line(0,-1){12}}
\put(112,50){\makebox(6,10){$\bullet$}}
\put(112,20){\makebox(6,10){$X$}} 
\put(115,25){\circle{6}} 
\put(115,55){\line(0,-1){27}}  
\put(122,20){\makebox(6,10){$\bullet$}}
\put(122, 5){\makebox(6,10){$V^{\dagger}$}} 
\put(125,10){\circle{6}} 
\put(125,25){\line(0,-1){12}}
\put(132,50){\makebox(6,10){$\bullet$}}
\put(132,20){\makebox(6,10){$X$}} 
\put(135,25){\circle{6}} 
\put(135,55){\line(0,-1){27}}  
\put(142,20){\makebox(6,10){$\bullet$}}
\put(142, 5){\makebox(6,10){$V$}} 
\put(145,10){\circle{6}} 
\put(145,25){\line(0,-1){12}}
\end{picture}
\end{center}
%
where $V$ is a unitary matrix given by 
\[
V=
\left(
  \begin{array}{cc}
    \frac{3+i}{4} & \frac{1-i}{4} \\
    \frac{1-i}{4} & \frac{3+i}{4}
  \end{array}
\right)
\quad \Longrightarrow \quad 
V^{4}=
\left(
  \begin{array}{cc}
    0 & 1 \\
    1 & 0 
  \end{array}
\right)
=\sigma_{1}.
\]

By the way, we have given the exact form of evolution operator for the four 
atoms Tavis--Cummings model \cite{Papers-2}, therefore we can in principle 
track the same line shown in this paper and it may be possible to get 
the controlled--controlled--controlled NOT gate or 
controlled--controlled--controlled unitary gates directly. 
However, such a calculation for the case becomes very complicated 
because we must treat $16\times 16$ matrices at each step of calculations. 
We will attempt it in the near future. 

It is not easy to extend our method to the general case because to obtain 
the exact form of evolution operator for the general Tavis--Cummings model 
(see (\ref{eq:hamiltonian-2})) is almost impossible. 

By the way, in \cite{FHKW} we have given an idea on construction of the 
controlled NOT gate of type (A) by making use of the following skillful method 
(see Figure 3).

\par \vspace{3mm} \noindent 
{\bf (i)}\ We move the third atom from the cavity. 
\par \noindent 
{\bf (ii)}\ We insert a photon in the cavity as two atoms interact with it 
and subject laser fields to the atoms, and next exchange the two atoms, 
which gives the controlled NOT operator as shown in the preceding section. 
\par \noindent 
{\bf (iii)}\ We return the third atom (outside the cavity) to the former 
position.

\begin{figure}
\begin{center}
\setlength{\unitlength}{1mm} 
\begin{picture}(150,40)(0,-20)
\bezier{200}(10,0)(0,10)(10,20)
\put(10,0){\line(0,1){20}}
\put(20,10){\circle*{3}}
\put(30,10){\circle*{3}}
\put(40,10){\circle*{3}}
\bezier{200}(50,0)(60,10)(50,20)
\put(50,0){\line(0,1){20}}
\put(65,10){\vector(1,0){20}}
\bezier{200}(100,0)(90,10)(100,20)
\put(100,0){\line(0,1){20}}
\put(110,10){\circle*{3}}
\put(120,10){\circle*{3}}
\put(130,22){\circle*{3}}
\bezier{200}(140,0)(150,10)(140,20)
\put(140,0){\line(0,1){20}}
\end{picture}
\end{center}
\vspace{-5mm}
\setlength{\unitlength}{1mm} 
\begin{picture}(150,20)(0,-30)
\put(125,0){\vector(0,-1){20}}
\end{picture}
\vspace{-5mm}
\begin{center}
\setlength{\unitlength}{1mm} 
\begin{picture}(150,40)(0,-20)
\bezier{200}(100,0)(90,10)(100,20)
\put(100,0){\line(0,1){20}}
\put(110,10){\circle*{3}}
\bezier{200}(110,-4)(112,-2)(110,0)
\bezier{200}(110,0)(108,2)(110,4)
\put(110,4){\line(0,1){2}}
\put(108.6,4){$\wedge$}
\put(120,10){\circle*{3}}
\bezier{200}(120,-4)(122,-2)(120,0)
\bezier{200}(120,0)(118,2)(120,4)
\put(120,4){\line(0,1){2}}
\put(118.6,4){$\wedge$}
\put(130,22){\circle*{3}}
\bezier{200}(140,0)(150,10)(140,20)
\put(140,0){\line(0,1){20}}
\put(90,10){\dashbox(60,0)}
\put(149,9){$>$}
\put(85,10){\vector(-1,0){20}}
%
%
\bezier{200}(10,0)(0,10)(10,20)
\put(10,0){\line(0,1){20}}
\put(20,18){\vector(0,-1){6.5}}
\put(20,10){\circle*{3}}
\put(30,18){\vector(0,-1){6.5}}
\put(30,10){\circle*{3}}
\put(40,22){\circle*{3}}
\put(15,18){\makebox(20,10)[c]{Exchange}}
\put(20,18){\line(1,0){10}}
\bezier{200}(50,0)(60,10)(50,20)
\put(50,0){\line(0,1){20}}
\end{picture}
\end{center}
\vspace{-5mm}
\setlength{\unitlength}{1mm} 
\begin{picture}(150,20)(0,-30)
\put(35,0){\vector(0,-1){20}}
\end{picture}
\vspace{-5mm}
\begin{center}
\setlength{\unitlength}{1mm} 
\begin{picture}(150,40)(0,-20)
\bezier{200}(10,0)(0,10)(10,20)
\put(10,0){\line(0,1){20}}
\put(20,10){\circle*{3}}
\put(30,10){\circle*{3}}
\put(40,10){\circle*{3}}
\bezier{200}(50,0)(60,10)(50,20)
\put(50,0){\line(0,1){20}}
\end{picture}
\end{center}
\vspace{-20mm}
\begin{center}
\caption{The process to construct the controlled NOT gate between the first 
atom and second one for the three atoms in a cavity}
\end{center}
\end{figure}
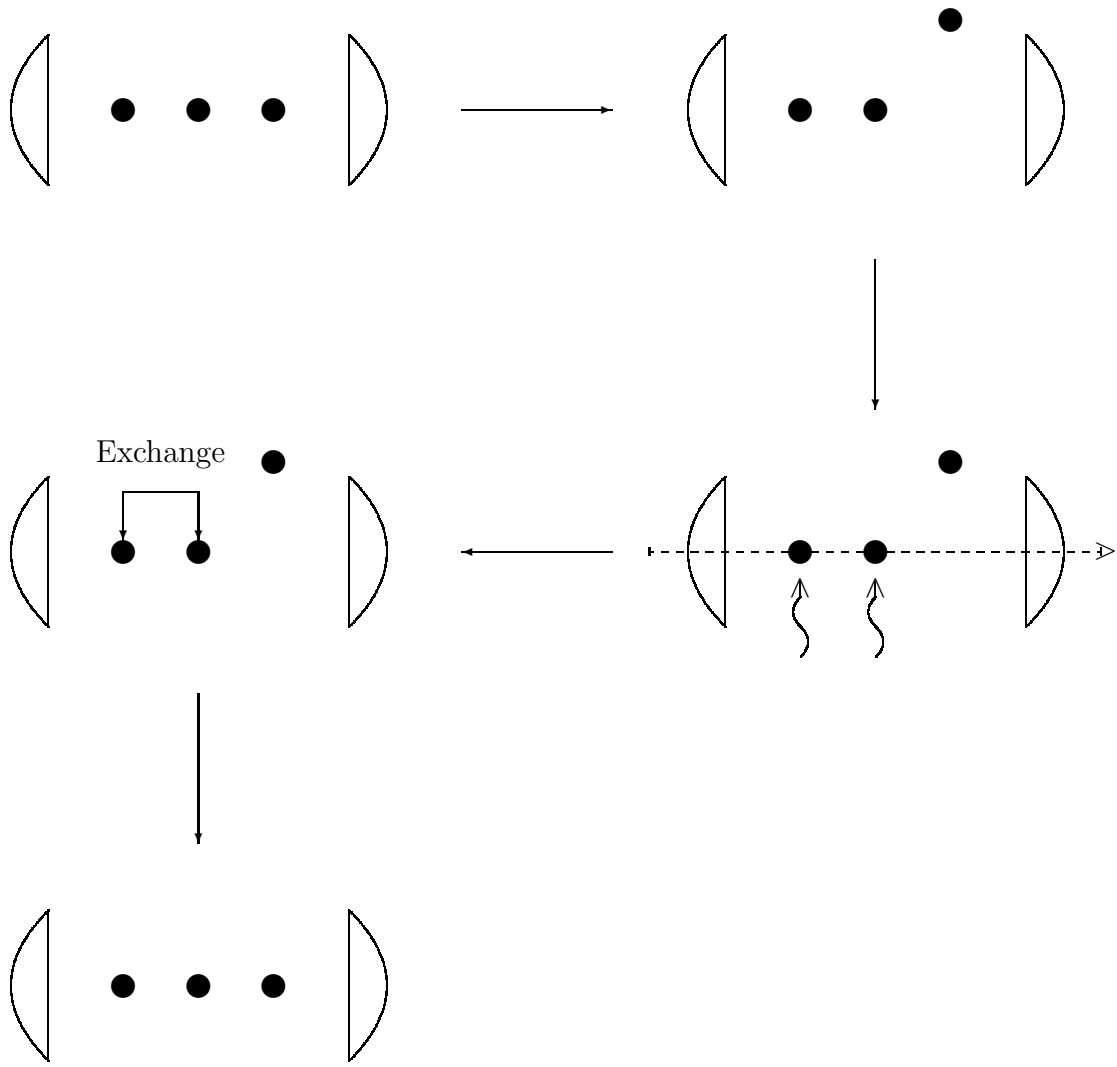
%

\vspace{5mm}
It is easy to generalize our idea to the case of $n$ atoms. 

One of important features of our model is that $n$ atoms are 
trapped {\bf linearly} in the cavity, which has both strong points and 
weak points. One of weak points is that a photon inserted in the cavity 
interacts with all atoms, so it is impossible to select atoms which 
interact with the photon. To improve this we presented the idea above.@

To perform a full quantum computation we need to construct (many) 
controlled--controlled NOT gates or controlled--controlled unitary ones 
for three atoms among $n$--atoms, see \cite{nine}; $\S$7. We believe that 
our idea is still available 
\footnote{We must estimate an influence of the 
``getting atoms (which are not our target) in and out" on the whole states 
space, which is however difficult in our model. For that we must 
add in (\ref{eq:hamiltonian-1}) further terms necessary to calculate it}. 

In principle, we can construct general quantum networks.

\section{Discussion}

In this paper we constructed the controlled NOT operator in the system of two 
qubits and controlled--controlled NOT operator in the system of three qubits 
in the quantum computation based on Cavity QED, which show that our system is 
universal. Therefore we can in principle perform a quantum computation. 
Our method is completely mathematical and we use several Rabi oscillations 
in a consistent manner.

\par \noindent 
{\bf 
We hope that some experimentalists will check whether our method works good or 
not.
}

\par \noindent 
See \cite{Brune et al} and their references for some experiments on Cavity QED 
(which may be related to our method).

We conclude this paper by making a comment (which is important at least 
to us). The Tavis--Cummings model is based on (only) two energy levels of 
atoms. However, an atom has in general infinitely many energy levels, 
so it might be natural to use this possibility. 
We are also studying a quantum computation based on multi--level systems of 
atoms (a qudit theory) \cite{Papers-4}. Therefore we would like to extend 
the Tavis--Cummings model based on two--levels to a model based on 
multi--levels. This is a very challenging task.

\vspace{5mm}
\noindent
{\it Acknowledgment.}
We wish to thank Tatsuo Suzuki and Shin'ichi Nojiri for their helpful comments 
and suggestions. 

\par \vspace{10mm}
\begin{center}
 \begin{Large}
   {\bf Appendix}
 \end{Large}
\end{center}

\par \vspace{5mm} \noindent 
{\bf \ \ One Qubit Operators by Classical Fields}

Let us make a brief review of theory without the radiation field, whose 
states space is only tensor product of two level systems of each atom. 
See the figure 5. 

\begin{figure}
\begin{center}
\setlength{\unitlength}{1mm} 
\begin{picture}(110,40)(0,-20)
\bezier{200}(20,0)(10,10)(20,20)
\put(20,0){\line(0,1){20}}
\put(30,10){\circle*{3}}
\bezier{200}(30,-4)(32,-2)(30,0)
\bezier{200}(30,0)(28,2)(30,4)
\put(30,4){\line(0,1){2}}
\put(28.6,4){$\wedge$}
\put(40,10){\circle*{3}}
\bezier{200}(40,-4)(42,-2)(40,0)
\bezier{200}(40,0)(38,2)(40,4)
\put(40,4){\line(0,1){2}}
\put(38.6,4){$\wedge$}
\put(50,10){\circle*{1}}
\put(60,10){\circle*{1}}
\put(70,10){\circle*{1}}
\put(50,1){\circle*{1}}
\put(60,1){\circle*{1}}
\put(70,1){\circle*{1}}
\put(80,10){\circle*{3}}
\bezier{200}(80,-4)(82,-2)(80,0)
\bezier{200}(80,0)(78,2)(80,4)
\put(80,4){\line(0,1){2}}
\put(78.6,4){$\wedge$}
\bezier{200}(90,0)(100,10)(90,20)
\put(90,0){\line(0,1){20}}
%
\end{picture}
\vspace{-15mm}
\caption{The $n$ atoms in the cavity without a photon (in Figure 1)}
\end{center}
\end{figure}
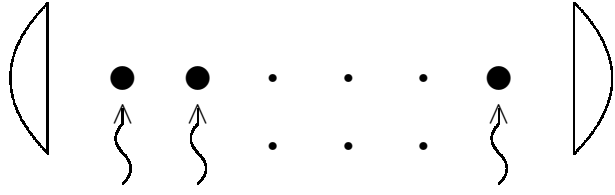

The Hamiltonian in this case is 
\begin{equation}
\label{eq:hamiltonian-special}
H=
\sum_{j=1}^{n}
\left\{
\frac{\Delta}{2}\sigma^{(3)}_{j}+
h_{j}
\left(\sigma^{(+)}_{j}\mbox{e}^{i(\Omega_{j}t+\phi_{j})}+
\sigma^{(-)}_{j}\mbox{e}^{-i(\Omega_{j}t+\phi_{j})} \right)
\right\}
\end{equation}
from (\ref{eq:hamiltonian-1}), where we have omitted the unit operator 
${\bf 1}$ for simplicity. 
Here let us remind the notation 
\[
M_{j}=1_{2}\otimes \cdots \otimes 1_{2}\otimes M \otimes 1_{2}\otimes 
\cdots \otimes 1_{2}\quad \mbox{for}\quad 1\leq j \leq n.
\]
Then 
\begin{eqnarray}
H
&=&
\sum_{j=1}^{n}
\left(
\begin{array}{cc}
 \frac{\Delta}{2} & h_{j}\mbox{e}^{i(\Omega_{j}t+\phi_{j})}     \\
 h_{j}\mbox{e}^{-i(\Omega_{j}t+\phi_{j})} & -\frac{\Delta}{2} 
\end{array}
\right)_{j}       \nonumber \\
&=&
\sum_{j=1}^{n}
\left\{
\left(
\begin{array}{cc}
 \mbox{e}^{i\frac{\Omega_{j}t+\phi_{j}}{2}}  &    \\
    & \mbox{e}^{-i\frac{\Omega_{j}t+\phi_{j}}{2}}
\end{array}
\right)
\left(
\begin{array}{cc}
 \frac{\Delta}{2} & h_{j}   \\
 h_{j} & -\frac{\Delta}{2} 
\end{array}
\right)
\left(
\begin{array}{cc}
 \mbox{e}^{-i\frac{\Omega_{j}t+\phi_{j}}{2}}  &   \\
    & \mbox{e}^{i\frac{\Omega_{j}t+\phi_{j}}{2}}
\end{array}
\right)
\right\}_{j}   \nonumber \\
&=&
\left(U_{1}\otimes \cdots \otimes U_{n}\right)
\sum_{j=1}^{n}
\left(
\begin{array}{cc}
 \frac{\Delta}{2} & h_{j}   \\
 h_{j} & -\frac{\Delta}{2} 
\end{array}
\right)_{j}
\left(U_{1}\otimes \cdots \otimes U_{n}\right)^{\dagger},
\end{eqnarray}
where 
\[
U_{j}=
\left(
\begin{array}{cc}
 \mbox{e}^{i\frac{\Omega_{j}t+\phi_{j}}{2}}  &    \\
    & \mbox{e}^{-i\frac{\Omega_{j}t+\phi_{j}}{2}}
\end{array}
\right).
\]

The wave function defined by $i\frac{d}{dt}\ket{\Psi}=H\ket{\Psi}$ with 
(\ref{eq:hamiltonian-special}) can be written as a tensor product 
\begin{equation}
\ket{\Psi}=\ket{\psi_{1}}\otimes \cdots \otimes \ket{\psi_{n}},
\end{equation}
so if we define 
\[
\ket{\tilde{\Psi}}\equiv 
\left(U_{1}\otimes \cdots \otimes U_{n}\right)^{\dagger}\ket{\Psi},
\]
then it is easy to see 
\[
i\frac{d}{dt}\ket{\tilde{\Psi}}
=
\sum_{j=1}^{n}
\left(
\begin{array}{cc}
 \frac{\Delta-\Omega_{j}}{2} & h_{j}   \\
 h_{j} & -\frac{\Delta-\Omega_{j}}{2} 
\end{array}
\right)_{j}
\ket{\tilde{\Psi}}.
\]
Th solution is easy to obtain 
\[
\ket{\tilde{\Psi}(t)}=
\bigotimes_{j=1}^{n}\mbox{exp}
\left\{
-it
\left(
\begin{array}{cc}
 \frac{\Delta-\Omega_{j}}{2} & h_{j}   \\
 h_{j} & -\frac{\Delta-\Omega_{j}}{2} 
\end{array}
\right)
\right\}
\ket{\tilde{\Psi}(0)}.
\]
Therefore, the solution that we are looking for is 
\begin{eqnarray}
\ket{\Psi(t)}
&=&\left(U_{1}\otimes \cdots \otimes U_{n}\right)\ket{\tilde{\Psi}(t)}
\nonumber \\
&=&\bigotimes_{j=1}^{n}
\left(
\begin{array}{cc}
 \mbox{e}^{i\frac{\Omega_{j}t+\phi_{j}}{2}}  &    \\
    & \mbox{e}^{-i\frac{\Omega_{j}t+\phi_{j}}{2}}
\end{array}
\right)
\mbox{exp}
\left\{-it
\left(
\begin{array}{cc}
 \frac{\Delta-\Omega_{j}}{2} & h_{j}   \\
 h_{j} & -\frac{\Delta-\Omega_{j}}{2} 
\end{array}
\right)
\right\}
\ket{{\Psi}(0)}.
\end{eqnarray}

Last we note that
\[
\mbox{exp}
\left\{-it
\left(
\begin{array}{cc}
 \frac{\theta}{2} & h  \\
 h & -\frac{\theta}{2} 
\end{array}
\right)
\right\}
=
\left(
\begin{array}{cc}
  x_{11} & x_{12}  \\
  x_{21} & x_{22}
\end{array}
\right)
\]
where 
\begin{eqnarray}
x_{11}&=&\mbox{cos}\left(t\sqrt{\frac{\theta^{2}}{4}+h^{2}}\right)
-i\frac{\theta}{2}
\frac{\mbox{sin}\left(t\sqrt{\frac{\theta^{2}}{4}+h^{2}}\right)}
     {\sqrt{\frac{\theta^{2}}{4}+h^{2}}}, \nonumber \\
x_{12}&=&x_{21}=-ih
\frac{\mbox{sin}\left(t\sqrt{\frac{\theta^{2}}{4}+h^{2}}\right)}
     {\sqrt{\frac{\theta^{2}}{4}+h^{2}}}, \nonumber \\
x_{22}&=&\mbox{cos}\left(t\sqrt{\frac{\theta^{2}}{4}+h^{2}}\right)
+i\frac{\theta}{2}
\frac{\mbox{sin}\left(t\sqrt{\frac{\theta^{2}}{4}+h^{2}}\right)}
     {\sqrt{\frac{\theta^{2}}{4}+h^{2}}}. \nonumber
\end{eqnarray}
We can always construct unitary operators in $U(2)$ at each atom by 
using Rabi osillations, see for example \cite{KF2}. 

As an example let us construct the Walsh--Hadamard operator $W$ which has 
been used in the text. We set
\[
V(t)=
\left(
\begin{array}{cc}
 \mbox{e}^{i\frac{\Omega t+\phi}{2}}  &    \\
    & \mbox{e}^{-i\frac{\Omega t+\phi}{2}}
\end{array}
\right)
\mbox{exp}
\left\{-it
\left(
\begin{array}{cc}
 \frac{\Delta-\Omega}{2} & h   \\
 h & -\frac{\Delta-\Omega}{2} 
\end{array}
\right)
\right\}
\]
and set the resonance condition $\Delta=\Omega$, then 
\begin{eqnarray*}
V(t)
&=&
\left(
\begin{array}{cc}
 \mbox{e}^{i\frac{\Delta t+\phi}{2}}  &    \\
    & \mbox{e}^{-i\frac{\Delta t+\phi}{2}}
\end{array}
\right)
\left(
\begin{array}{cc}
 \mbox{cos}(ht) & -i\mbox{sin}(ht) \\
 -i\mbox{sin}(ht) & \mbox{cos}(ht)
\end{array}
\right)     \\
&=&
\mbox{e}^{i\frac{\Delta t+\phi}{2}}
\left(
\begin{array}{cc}
 1  &    \\
    & \mbox{e}^{-i(\Delta t+\phi)}
\end{array}
\right)
\left(
\begin{array}{cc}
 \mbox{cos}(ht) & -i\mbox{sin}(ht) \\
 -i\mbox{sin}(ht) & \mbox{cos}(ht)
\end{array}
\right)
\end{eqnarray*}
Now we again set
\begin{equation}
V(t)=
\left(
\begin{array}{cc}
 1  &    \\
    & \mbox{e}^{-i(\Delta t+\phi)}
\end{array}
\right)
\left(
\begin{array}{cc}
 \mbox{cos}(ht) & -i\mbox{sin}(ht) \\
 -i\mbox{sin}(ht) & \mbox{cos}(ht)
\end{array}
\right)
\end{equation}
for simplicity because there is no interest in the overall phase factor 
\footnote{In the Hamiltonian (\ref{eq:hamiltonian-1}) a constant term which 
makes a overall phase factor has been removed}. 
On the other hand, when we don't subject a laser field to the atom ($h=0$) 
we have
\begin{equation}
V_{0}(t)=
\left(
\begin{array}{cc}
 1  &    \\
    & \mbox{e}^{-i\Delta t}
\end{array}
\right).
\end{equation}
By using $V(t)$ and $V_{0}(t)$ we construct $W$ in the following 
(\cite{KF2}). First we set 
\[
V(t_{f},t_{i})=V(t_{f}-t_{i}),\quad 
V_{0}(t_{f},t_{i})=V_{0}(t_{f}-t_{i})\quad \mbox{for}\quad t_{f}\ >\ t_{i}. 
\]
A sequence of operators to construct the Walsh--Hadamard gate is given as 
follows : 

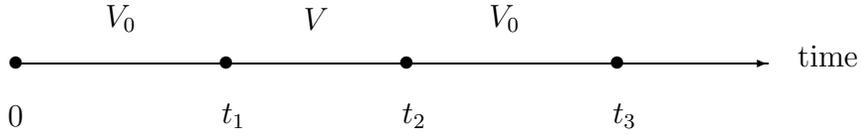
\begin{center}
\setlength{\unitlength}{1mm} 
\begin{picture}(140,30)
\put(10,15){\vector(1,0){100}}
\put( 7,10){\makebox(6,10){$\bullet$}}
\put( 7, 3){\makebox(6,10){$0$}}
\put(21,17){\makebox(6, 8){$V_{0}$}}
\put(35,10){\makebox(6,10){$\bullet$}}
\put(36, 3){\makebox(6,10){$t_{1}$}}
\put(47,17){\makebox(6, 8){$V$}}
\put(59,10){\makebox(6,10){$\bullet$}}
\put(60, 3){\makebox(6,10){$t_{2}$}}
\put(72,17){\makebox(6, 8){$V_{0}$}}
\put(87,10){\makebox(6,10){$\bullet$}}
\put(88, 3){\makebox(6,10){$t_{3}$}}
\put(115,11){\makebox(6,10){time}}
\end{picture}
\end{center}

For $V_{0}(t_{1},0)$, 
$V_{0}(t_{3},t_{2})$ with $t_{1}=3\pi/2\Delta$ and $t_{3}-t_{2}=3\pi/2\Delta$ 
\[
V_{0}(t_{1},0)
=
\left(
  \begin{array}{cc}
    1& 0 \\
    0& i
  \end{array}
\right), \quad 
V_{0}(t_{3},t_{2})
=
\left(
  \begin{array}{cc}
    1& 0 \\
    0& i
  \end{array}
\right)
\]
and $V(t_{2},t_{1})$ with $t_{2}-t_{1}=\pi/4h$
\[
V(t_{2},t_{1})
=
\left(
  \begin{array}{cc}
    1&         \\
     & \mbox{e}^{-i\{\Delta(t_{2}-t_{1})+\phi\}}
  \end{array}
\right)
\left(
  \begin{array}{cc}
    \frac{1}{\sqrt{2}} & \frac{-i}{\sqrt{2}}  \\
    \frac{-i}{\sqrt{2}}& \frac{1}{\sqrt{2}} 
  \end{array}
\right), 
\]
we have 
\[
V_{0}(t_{3},t_{2})V(t_{2},t_{1})V_{0}(t_{1},0)
=
\left(
  \begin{array}{cc}
    1&         \\
     & \mbox{e}^{-i\{\Delta(t_{2}-t_{1})+\phi\}}
  \end{array}
\right)
\left(
  \begin{array}{cc}
    \frac{1}{\sqrt{2}}& \frac{1}{\sqrt{2}}  \\
    \frac{1}{\sqrt{2}}& -\frac{1}{\sqrt{2}} 
  \end{array}
\right).
\]
By choosing the phase $\phi$ 
as $\mbox{e}^{-i\{\Delta(t_{2}-t_{1})+\phi\}}=1$ 
we finally obtain 
\begin{equation}
V_{0}(t_{3},t_{2})V(t_{2},t_{1})V_{0}(t_{1},0)
=
\frac{1}{\sqrt{2}}
\left(
  \begin{array}{cc}
    1& 1  \\
    1& -1
  \end{array}
\right)=W. 
\end{equation}

\vspace{10mm}


\begin{thebibliography}{99}
%
\bibitem{Books}L. Allen and J. H. Eberly : 
\newblock Optical Resonance and Two--Level Atoms, 
\newblock Wiley, New York, 1975.\ 
%
P. Meystre and M. Sargent III : 
\newblock Elements of Quantum Optics (third edition), 
\newblock Springer--Verlag, 1990.\ 
%
Claude Cohen--Tannoudji, J. Dupont--Roc and G. Grynberg : 
\newblock Atom--Photon Interactions ; Basic Processes and Applications, 
\newblock Wiley, New York, 1998. 
%
\bibitem{decoherence}W. H. Zurek : 
\newblock Decoherence and the transition from quantum to classical--REVISITED, 
\newblock quant-ph/0306072.\ 
%
M. Frasca : 
\newblock Dynamical decoherence in a cavity with a large number of two-level 
atoms, 
\newblock J. Phys. B : At. Mol. Opt. Phys. 37(2004) 1273, 
\newblock quant-ph/0310117. 
%
\bibitem{FHKW}K. Fujii, K. Higashida, R. Kato and Y. Wada : 
\newblock Cavity QED and Quantum Computation in the Weak Coupling Regime, 
\newblock J. Opt. B: Quantum and Semiclass. Opt, 6(2004) 502, 
\newblock quant-ph/0407014. 
%
\bibitem{Papers-1}E. T. Jaynes and F. W. Cummings : 
\newblock Comparison of Quantum and Semiclassical Radiation Theories with 
Applications to the Beam Maser, 
\newblock Proc. IEEE 51(1963), 89.\ 
%
M. Tavis and F. W. Cummings : 
\newblock Exact Solution for an N--Molecule--Radiation--Field Hamiltonian, 
\newblock Phys. Rev. 170(1968), 379.
%
\bibitem{nine} A. Barenco, C. H. Bennett, R. Cleve, D. P. Vincenzo, N. 
Margolus, P. Shor, T. Sleator, J. Smolin and H. Weinfurter : 
Elementary gates for quantum computation, Phys. Rev. A 52, 3457, 1995, 
quant-ph/9503016. 
%
\bibitem{KF1}K. Fujii : 
\newblock Introduction to Grassmann Manifolds and Quantum Computation, 
\newblock J. Applied Math, 2(2002), 371, 
\newblock quant-ph/0103011. 
%
\bibitem{Papers-2}K. Fujii, K. Higashida, R. Kato and Y. Wada : 
\newblock Explicit Form of Solution of Two Atoms Tavis--Cummings Model, 
\newblock to appear in The Bulletin of Yokohama City University, vol. 54, 
2005, 
\newblock quant-ph/0403008.\ 
%
K. Fujii, K. Higashida, R. Kato, T. Suzuki and Y. Wada : 
\newblock Explicit Form of the Evolution Operator of Tavis--Cummings Model :
Three and Four Atoms Cases, Int. J. Geom. Methods Mod. Phys, Vol.1, 
No.6(2004), 721, 
\newblock quant-ph/0404034.
%
\bibitem{Papers-3}M. Orszag, R. Ramirez, J. C. Retamal and C. Saavedra : 
\newblock Quantum cooperative effects in a micromaser, 
\newblock Phys. Rev. A 49 (1994), 2933.\ 
%
M. S. Kim, J. Lee, D. Ahn and P. L. Knight : 
\newblock Entanglement induced by a single-mode heat environment, 
\newblock Phys. Rev. A 65 (2002), 040101, 
\newblock quant-ph/0109052.\ 
%
C. Genes, P. R. Berman and a. G. Rojo : 
\newblock Spin squeezing via atom -- cavity field coupling, 
\newblock quant-ph/0306205. 
%
\bibitem{BB}J-L. Brylinski and R. Brylinski : 
\newblock Universal Quantum Gates, 
\newblock quant-ph/0108062. 
%
\bibitem{Dirac}P. A. M. Dirac : 
\newblock The Principles of Quantum Mechanics, 
\newblock Oxford University Press, 1958 ; {\bf 55} Permutations as 
dynamical variables. 
%
\bibitem{KF2}K. Fujii : 
\newblock Quantum Optical Construction of Generalized Pauli and 
Walsh--Hadamard Matrices in Three Level Systems, 
\newblock quant-ph/0309132. 
%
\bibitem{Papers-4}K. Fujii : 
\newblock Exchange Gate on the Qudit Space and Fock Space, 
\newblock J. Opt. B : Quantum Semiclass. Opt, 5(2003), S613, 
\newblock quant-ph/0207002.\ 
%
K. Fujii, K. Higashida, R. Kato and Y. Wada : 
\newblock N Level System with RWA and Analytical Solutions Revisited, 
\newblock quant-ph/0307066.\ 
%
K. Fujii, K. Higashida, R. Kato and Y. Wada : 
\newblock A Rabi Oscillation in Four and Five Level Systems, 
\newblock quant-ph/0312060.\ 
%
K. Funahashi : 
\newblock Explicit Construction of Controlled--U and Unitary Transformation 
in Two--Qudit, 
\newblock quant-ph/0304078. 
%
\bibitem{Brune et al}M. Brune, F. Schmidt-Kaler, A. Maali, J. Dreyer, 
E. Hagley, J. M. Raimond and S. Haroche : 
\newblock Quantum Rabi Oscillation : A Direct Test of Field Quantization 
in a Cavity, 
\newblock Phys. Rev. Lett, 76(1996), 1800.\ 
%
X. Maitre, E. Hagley, G. Nogues, C. W. Wunderlich, P. Goy, M. Brune, 
J. M. Raimond and S. Haroche : 
\newblock Quantum Memory with a Single Photon in a Cavity, 
\newblock Phys. Rev. Lett, 79(1997), 769. 
%
\end{thebibliography}
\end{document}